\documentclass[12pt,twoside]{article}
\usepackage{epsfig}
\newlength{\dinwidth}                                                           
\newlength{\dinmargin}                                                          
\usepackage{amssymb}

\vspace*{0.1cm}

\begin{document}

\begin{center}
\LARGE{The Superconducting TESLA Cavities}
\end{center}

\frenchspacing

\bigskip

\begin{center}
\large{Dedicated to the memory of Bj{\o}rn H. Wiik}
\end{center}

\vspace{0.2cm}

\begin{raggedright}
\noindent
B.~Aune$^{1}$, R.~Bandelmann$^4$, D.~Bloess$^2$, B.~Bonin$^{1,11}$, A.~Bosotti$^7$, M.~Champion$^5$,
C.~Crawford$^5$, G. Deppe$^4$, B. Dwersteg$^4$, D.A. Edwards$^{4,5}$, H.T Edwards$^{4,5}$, 
M. Ferrario$^6$, M. Fouaidy$^9$, P.-D. Gall$^4$, A.~Gamp$^4$, A.~G\"ossel$^4$, J.~Graber$^{4,12}$,  D.~Hubert$^4$,
M.~H\"uning$^4$, M.~Juillard$^1$, T.~Junquera$^9$, H.~Kaiser$^4$, G.~Kreps$^4$, M.~Kuchnir$^5$, R.~Lange$^4$,
M.~Leenen$^4$, M.~Liepe$^4$, L.~Lilje$^4$, A.~Matheisen$^4$, W.-D.~M\"oller$^4$, A.~Mosnier$^1$,
H.~Padamsee$^3$, C.~Pagani$^7$, M.~Pekeler$^{4,10}$, H.-B.~Peters$^4$, O.~Peters$^4$, D.~Proch$^4$,
K.~Rehlich$^4$, D.~Reschke$^4$, H.~Safa$^1$, T.~Schilcher$^{4,13}$, P.~Schm\"user$^8$, J.~Sekutowicz$^4$, S.~Simrock$^4$, W.~Singer$^4$, M.~Tigner$^3$, D.~Trines$^4$, K.~Twarowski$^4$, G.~Weichert$^4$, J.~Weisend$^{4,14}$, J.~Wojtkiewicz$^4$, S.~Wolff$^4$ and K.~Zapfe$^4$
\end{raggedright}

\vspace{0.2cm}

\begin{flushleft}
$ ^{1}$ CEA Saclay, Gif-sur-Yvette, France \\ 
$ ^{2}$ CERN, Geneve, Switzerland \\
$ ^{3}$ Cornell University, Ithaca, New York, USA \\
$ ^{4}$ Deutsches Elektronen Synchrotron, Hamburg, Germany \\
$ ^{5}$ FNAL, Batavia, Illinois, USA \\
$ ^{6}$ INFN Frascati, Frascati, Italy \\
$ ^{7}$ INFN Milano, Milano, Italy \\
$ ^{8}$ II. Institut f\"ur Experimentalphysik, Universit\"at Hamburg, Germany \\
$ ^{9}$ I.P.N. Orsay, Orsay, France \\
$ ^{10}$ Now at ACCEL Instruments GmbH, Bergisch-Gladbach, Germany \\
$ ^{11}$ Now at CEA/IPSN, Fontenay, France \\
$ ^{12}$ Now at Center for Advanced Biotechnology, Boston University, USA \\
$ ^{13}$ Now at Paul Scherrer Institut, Villigen, Switzerland \\
$ ^{14}$ Now at SLAC, Stanford, California, USA 
\end{flushleft}

\vspace{0.4cm}

\begin{abstract}
The conceptional design of the proposed linear electron-positron collider TESLA is based on 9-cell 1.3 GHz superconducting niobium cavities with an accelerating gradient of $E_{acc}\ge 25$ MV/m at a quality factor $Q_0 \ge 5 \cdot 10^9$. The design goal for the cavities of the TESLA Test Facility (TTF) linac was set to the more moderate value of $E_{acc}\ge 15$ MV/m. In a first series of 27 industrially produced TTF cavities the average gradient at $Q_0=5 \cdot 10^9$ was measured to be $20.1\pm6.2$ MV/m, excluding a few cavities suffering from serious fabrication or material defects. In the second production of 24 TTF cavities additional quality control measures were introduced, in particular an eddy-current scan to eliminate niobium sheets with foreign material inclusions and stringent prescriptions for carrying out the electron-beam welds. The average gradient of these cavities at $Q_0 = 5 \cdot 10^9$ amounts to $25.0\pm3.2$ MV/m with the exception of one cavity suffering from a weld defect. Hence only a moderate improvement in production and preparation techniques will be needed to meet the ambitious TESLA goal with an adequate safety margin. In this paper we present a detailed description of the design, fabrication and preparation of the TESLA Test Facility cavities and their associated components and report on cavity performance in test cryostats and  with electron beam in the TTF linac. The ongoing R\&D towards  higher gradients is briefly addressed.
\end{abstract}

\clearpage

\section{Introduction}

In the past 30 years electron-positron collisions have played a central role in the discovery and detailed investigation of new elementary particles and their interactions. The highly successful Standard Model of the unified electromagnetic and weak interactions and Quantum Chromodynamics, the quantum field theory of quark-gluon interactions, are to a large extent based on the precise data collected at electron-positron colliders. Important questions still remain to be answered, in particular the origin of the masses of field quanta and particles -- within the Standard Model explained in terms of the so-called Higgs mechanism -- and the existence or nonexistence of supersymmetric particles which appear to be a necessary ingredient of any quantum field theory attempting to unify all four forces known in Nature: the  gravitational, weak, electromagnetic and strong forces. There is  general agreement in the high energy physics community that in addition to the Large Hadron Collider under construction at CERN a lepton collider will be needed to address these fundamental issues. 

Electron-positron interactions in the center-of-mass (cm) energy range from 200 GeV to more than a TeV can no longer be realized in a circular machine like LEP since the $E^4$ dependence of the synchrotron radiation loss would lead to prohibitive operating costs. Instead the linear collider concept must be employed. This new principle was successfully demonstrated with the Stanford Linear Collider SLC providing a cm energy of more than 90 GeV. Worldwide there are different design options towards the next generation of linear colliders in the several 100 GeV to TeV regime. Two main routes are followed: colliders equipped with normal-conducting (nc) cavities  (NLC, JLC, VLEPP and CLIC) or with superconducting (sc) cavities (TESLA). The normal-conducting designs are based on high frequency structures (6 to 30 GHz) while the superconducting TESLA collider employs the comparatively low frequency of 1.3 GHz. The high conversion efficiency from primary electric power to beam power (about 20\%) in combination with the small beam emittance growth in low-frequency accelerating structures makes the superconducting option an ideal choice for a high-luminosity collider.

The first international TESLA workshop was held in 1990 \cite{TESLA90}. 
At that time superconducting rf cavities in particle accelerators were usually operated in the 5 MV/m regime. Such low gradients, together with the high cost of cryogenic equipment, would have made a superconducting linear electron-positron collider totally non-competitive with the normal-conducting colliders proposed in the USA and Japan. The TESLA collaboration, formally established in 1994 with the aim of developing a 500 GeV center-of-mass energy superconducting linear collider, set out with the ambitious goal of increasing the cost effectiveness of the superconducting option by more than an order of magnitude: firstly, by raising the accelerating gradient by a factor of five from 5 to 25~MV/m, and secondly, by reducing the cost per unit length of the linac by using economical cavity production methods and a greatly simplified cryostat design. Important progress has been achieved in both directions; in particular the gradient of 25 MV/m is essentially in hand, as will be shown below. To allow for a gradual improvement in the course of the cavity R\&D program, a more moderate goal of 15 MV/m was set for the TESLA Test Facility (TTF) linac \cite{TTF_LINAC}. 

The TESLA cavities are quite similar in their layout to the 5-cell 1.5 GHz cavities of the electron accelerator CEBAF in Newport News (Virginia, USA) which were made by an industrial company \cite{cebaf}. These cavities exceeded the specified gradient of 5 MV/m considerably and hence offered the potential for further improvement. 
While the CEBAF cavity fabrication methods were adopted for TTF without major modifications, important new steps were introduced in the cavity preparation:
\begin{itemize}
\item	chemical removal of a thicker surface layer
\item	a 1400$^\circ$C annealing with titanium getter to improve the Nb heat conductivity and to homogenize the material
\item	rinsing with ultrapure water at high pressure (100 bar) to remove surface contaminants
\item	destruction of field emitters by a technique called High Power Processing (HPP).
\end{itemize}
The application of these techniques, combined with extremely careful handling of the cavities in a clean-room environment, has led to a significant increase in accelerating field.

The TESLA Test Facility (TTF) has been set up at DESY to provide the necessary infra-structure for the chemical and thermal treatment, clean-room assembly and testing of industrially produced multicell cavities. In addition a 500~MeV electron linac is being built as a test bed for the performance of the sc accelerating structures with an electron beam of high bunch charge. At present more than 30 institutes from Armenia, P.R. China, Finland, France, Germany, Italy, Poland, Russia and USA participate in the TESLA Collaboration and contribute to TTF. 

The low frequency of 1.3~GHz permits the acceleration of long trains of particle bunches with very low emittance making a superconducting linac an ideal driver of a free electron laser (FEL) in the vacuum ultraviolet and X-ray regimes. For this reason the TTF linac has recently been equipped with undulator magnets, and its energy will be upgraded to 1 GeV in the coming years to provide an FEL user facility in the nanometer wavelength range. An X-ray FEL facility with wavelengths below 1~{\AA} is an integral part of the TESLA collider project \cite{design_rep}. 

The present paper is organized as follows. Sect.~\ref{prop} is devoted to the basics of rf superconductivity and the properties and limitations of superconducting (sc) cavities for particle acceleration. The design of the TESLA cavities and the auxiliary equipment is presented in Sect.~\ref{design}. The fabrication and preparation steps of the cavities are described in Sect.~\ref{fabrication}. The test results obtained on all TTF cavities are presented in Sect.~\ref{results}, together with a discussion of errors and limitations encountered during cavity production at industry, and the quality control measures taken. The rf control and the cavity performance with electron beam in the TTF linac are described in Sect.~\ref{rfcontrol}. A summary and outlook is given in Sect.~\ref{highgrad} where also the ongoing research towards higher gradients is shortly addressed.

\section{Basics of RF Superconductivity and Properties of Superconducting Cavities for Particle Acceleration  \label{prop}}

\subsection{Basic principles of rf superconductivity and choice of superconductor}

The existing large scale applications of superconductors in accelerators are twofold, in magnets and in accelerating cavities. While there are some common requirements like the demand for as high a critical temperature as possible\footnote{The High-$T_c$ ceramic superconductors have not yet found widespread application in magnets mainly due to technical difficulties in cable production and coil winding. Cavities with High-$T_c$ sputter coatings on copper have shown much inferior performance in comparison to niobium cavities.} there are also characteristic differences. In magnets operated with a dc or a low-frequency ac current the so-called ``hard'' superconductors are needed featuring high upper critical magnetic fields (15--20 T) and strong flux pinning in order to obtain high critical current density; such properties can only be achieved using alloys like niobium-titanium or niobium-tin. In microwave applications the limitation of the superconductor is not given by the upper critical field but rather by the so-called ``superheating field'' which is well below 1~T for all known superconductors. Moreover, strong flux pinning appears undesirable in microwave cavities as it is coupled with hysteretic losses. Hence a ``soft'' superconductor must be used and pure niobium is still the best candidate  although its critical temperature is only 9.2 K and the superheating field about 240 mT. Niobium-tin (Nb$_3$Sn) looks more favorable at first sight since it has a higher critical temperature of 18 K and a superheating field of 400 mT; however, the gradients achieved in Nb$_3$Sn coated single-cell copper cavities were below 15 MV/m, probably due to grain boundary effects in the Nb$_3$Sn layer \cite{wupp}. For these reasons the TESLA collaboration decided to use niobium as the superconducting material, as in all other large scale installations of sc cavities. Here two alternatives exist: the cavities are fabricated from solid niobium sheets or a thin niobium layer is sputtered onto the inner surface of a copper cavity. Both approaches have been successfully applied, the former at Cornell (CESR), KEK (TRISTAN), DESY (PETRA, HERA), Darmstadt (SDALINAC), Jefferson Lab (CEBAF) and other laboratories, the latter in particular at CERN in the electron-positron storage ring LEP. From the test results on existing cavities the solid-niobium approach promised higher accelerating gradients, hence it was adopted as the baseline for the TTF cavity R\&D program.

\subsubsection{Surface resistance}

In contrast to the dc case superconductors are not free from energy dissipation in microwave fields. The reason is that the  radio frequency (rf) magnetic field penetrates a  thin surface layer and induces oscillations of the electrons which are not bound in Cooper pairs. The number of these ``free electrons'' drops exponentially with temperature. According to the Bardeen-Cooper-Schrieffer (BCS) theory of superconductivity the surface resistance in the range $T<T_c/2$ is given by the expression

\begin{equation}
R_{BCS}\propto \frac {\omega^2} {T}\,
\exp(-1.76 \, T_c /T)      
\end{equation}    
where $f= \omega /2 \pi $
is the microwave frequency. In the two-fluid model of superconductors one can derive a refined expression for the surface resistance \cite{bonin, weingarten}
\begin{equation}    
R_{\rm BCS}= \frac{C}{T} \,
\omega^2 \, \sigma_n \, \Lambda^3 \, \exp(-1.76 \, T_c /T) \; .
\end{equation}    

\begin{figure}
\begin{center}
\epsfig{figure=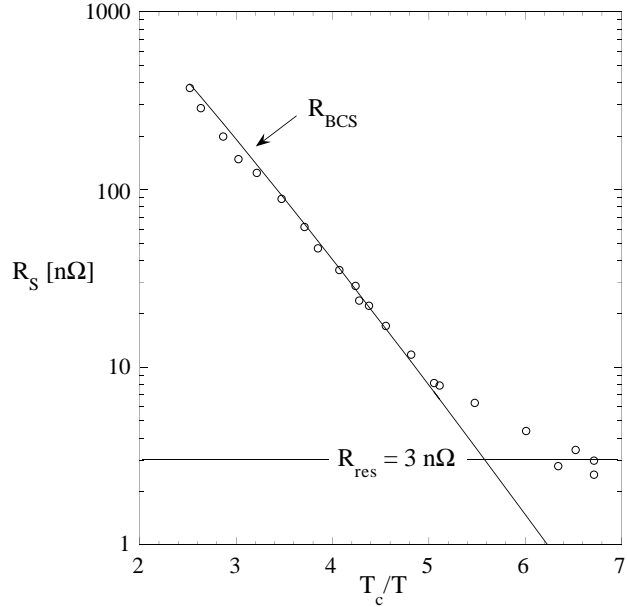,scale=0.48}
\caption{\label{rbcs} 
The surface resistance of a 9-cell TESLA cavity plotted as a 
function of $T_c/T$. The residual resistance of 3~n$\Omega$ corresponds to a quality factor $Q_0 = 10^{11}$.}
\end{center}
\end{figure}

\noindent
Here $C$ is a constant, $\sigma_n$ the normal-state
conductivity of the material and $\Lambda$ an effective penetration depth, given by
$$\Lambda=\lambda_L \sqrt {1+\xi_0/ \ell} \; .$$
$\lambda_L$ is the London penetration depth, $\xi_0$ the coherence length and $\ell$ the mean free path of the unpaired electrons. The fact that $\sigma_n$  is proportional to the mean free path
$\ell$ leads to the surprising conclusion that the surface resistance does not assume its minimum value when the superconductor is as pure as possible  
($\ell \gg \xi_0$) but rather in the range  $\ell \approx \xi_0$. For niobium the BCS surface resistance at 1.3~GHz amounts to about 800~n$\Omega$ at 4.2~K and drops to 15~n$\Omega$ at 2~K; see Fig.~\ref{rbcs}. The exponential temperature dependence is the reason why operation at 1.8--2 K  is essential for achieving high accelerating gradients in combination with very high quality factors. Superfluid helium is an excellent coolant owing to its high heat conductivity.  

In addition to the BCS term there is a  residual resistance $ R_{\rm res}$ caused by impurities, frozen-in magnetic flux or lattice distortions. This term is temperature independent and amounts to a few n$\Omega$ for very pure niobium but may readily increase if the surface is contaminated. 

\subsubsection{Heat conduction in niobium}

\begin{figure}
\begin{center}
\epsfig{figure=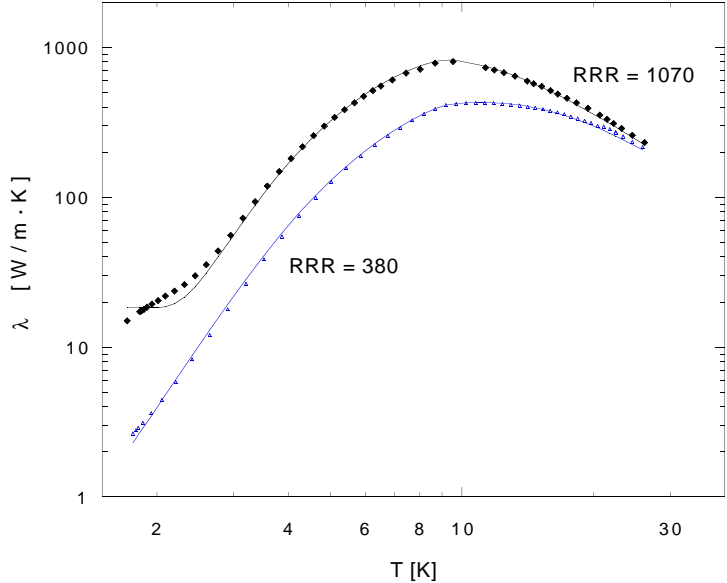,scale=0.6}
\caption{\label{heatconduct} Measured heat conductivity in 
niobium as a function of temperature \cite{schilcher}. 
Continuous curves: parametrization by B. Bonin, using the $RRR$ and the average grain size as input parameters \cite{bonin2}. These data do not show an enhancement at 2 K (the so-called ``phonon peak'') which was observed in some earlier experiments \cite{wupp}.}
\end{center}
\end{figure}

The heat produced at the inner cavity surface has to be guided through the cavity wall to the superfluid helium bath. Two quantities characterize the heat flow: the thermal conductivity of the bulk niobium and the temperature drop at the niobium-helium interface caused by the Kapitza resistance. For niobium with a residual resistivity ratio\footnote{$RRR$ is defined as the ratio of the resistivities at room temperature and at liquid helium temperature. The low temperature resistivity is either measured just above $T_c$ or at 4.2~K, applying a magnetic field to assure the normal state.} $RRR=500$ the two contributions to the temperature rise at the inner cavity surface are about equal. The thermal conductivity of niobium at cryogenic temperatures scales approximately with the $RRR$, a rule of thumb being
$$\lambda(\rm {4.2 K}) \approx 0.25 \cdot {\it RRR} \hspace{5mm} [\rm{W/(m \cdot K)}].$$
However, $\lambda$  is strongly temperature dependent and drops by about an order of magnitude when lowering the temperature to 2~K, as shown in Fig.~\ref{heatconduct}.

Impurities influence the $RRR$ and the thermal conductivity of niobium. Bulk niobium is contaminated by interstitial (mostly hydrogen, carbon, nitrogen, oxygen) and metallic impurities (mostly tantalum). The resulting $RRR$ can be calculated by summing the individual contributions \cite{Schulze}
\begin {equation}
RRR = \left( \sum_i {f_i/r_i} \right)^{-1}
\end {equation}
where the $f_i$ denote the fractional contents of impurity $i$ (measured in wt. ppm) and the $r_i$ the corresponding resistivity coefficients which are listed in Table~\ref{RRRtable}. 

\begin{table}
\begin{center}
\caption{\label{RRRtable} Resistivity coefficients of common impurities in Nb \cite{Schulze}}
\begin{tabular}{|c|c c c c c|}
\hline
Impurity atom $i$ & N & O & C & H & Ta \\
\hline
$r_i$ in $10^{4}$ wt. ppm & 0.44 & 0.58 & 0.47 & 0.36 & 111  \\
\hline
\end{tabular}
\end{center}
\end{table}

A good thermal conductivity is the main motivation for using high purity niobium with $RRR \approx 300$ as the material for cavity production. The $RRR$ may be further improved by post-purification of the entire cavity (see Sect.~\ref{results}). 

The Kapitza conductance depends on temperature and surface conditions. For pure niobium in contact with superfluid helium at 2 K it amounts to about $6000$\,W/(m$^2$K) \cite{kapitza}.  

\subsubsection{Influence of magnetic fields \label{shielding}}

\noindent
{\it Superheating field}. Superconductivity breaks down when the rf magnetic field exceeds the critical field of the superconductor. In the high frequency case the so-called ``superheating field'' is relevant which for niobium is about 20$\%$ higher than the thermodynamical critical field of 200\,mT \cite{matricon,hays}. 

\noindent
{\it Trapped magnetic flux}. Niobium is in principle a soft type II superconductor without flux pinning. In practice, however, weak magnetic dc fields are not expelled upon cooldown but remain trapped in the niobium. Each flux line contains a normal-conducting core whose area is roughly $\pi \xi_0^2$. The coherence length $\xi_0$ amounts to 40 nm in Nb. Trapped magnetic dc flux therefore results in a surface resistance \cite{bonin}
\begin{equation}    
R_{\rm mag}=(B_{\rm ext}/2 B_{c2})R_n  
\end{equation}
where $B_{ext}$ is the externally applied field, $B_{c2}$ the upper critical field and $R_n$ the surface resistance in the normal state\footnote{Benvenuti {\it et al.} \cite{Benv} attribute the magnetic surface resistance in niobium sputter layers to flux flow.}. At 1.3 GHz the surface resistance caused by trapped flux amounts to 3.5 n$\Omega /\mu$T for niobium. Cavities which are not shielded from the Earth's magnetic field are therefore limited to $Q_0$ values below $10^9$. 

\subsection{Advantages and limitations of superconducting cavities}
The fundamental advantage of superconducting cavities is the extremely low surface resistance of about 10 n$\Omega$ at 2 K. The typical quality factors of normal conducting cavities are 10$^4$--10$^5$ while for sc cavities they may exceed $10^{10}$, thereby reducing the rf losses by 5 to 6 orders of magnitude. In spite of the low efficiency of refrigeration there are considerable savings in primary electric power. Only a tiny fraction of the incident rf power is dissipated in the cavity walls, the lion's share is either transferred to the beam or reflected into a load. 

The physical limitation of a sc resonator is given by the requirement that the rf magnetic field at the inner surface has to stay below the superheating field of the superconductor (200--240 mT for niobium). For the TESLA cavities this implies a maximum accelerating field of  50--60 MV/m. In principle the quality factor should stay roughly constant when approaching this fundamental superconductor limit but in practice  the ``excitation curve'' $Q_0=Q_0(E_{acc})$ ends at considerably lower values, often accompanied with a strong decrease of $Q_0$ towards the highest gradient reached in the cavity. The main reasons for the performance degradation are excessive heating at impurities on the inner surface, field emission of electrons and multipacting\footnote{ ``Multipacting'' is a commonly used abbreviation for ``multiple impacting'' and designates the resonant multiplication of field emitted electrons which gain energy in the rf electromagnetic field and impact on the cavity surface where they induce secondary electron emission.}. 

\subsubsection{Thermal instability and field emission}

One basic limitation of the maximum field in a superconducting cavity is thermal instability. Temperature mapping at the outer cavity wall usually reveals that the heating by rf losses is not uniform over the whole surface but that certain spots exhibit larger temperature rises, often beyond the critical temperature of the superconductor. Hence the cavity becomes partially normal-conducting, associated with strongly enhanced power dissipation. Because of the exponential increase of surface resistance with temperature this may result in a run-away effect and eventually a quench of the entire cavity. 
Analytical models as well as numerical simulations are available to describe such an avalanche effect. Input parameters are the thermal conductivity of the superconductor, the size and resistance of the normal conducting spot and the Kapitza resistance. The tolerable defect size depends on the $RRR$ of the material and the desired field level. As a typical number, the diameter of a normal-conducting spot must exceed 50~$\mu$m to be able to initiate a thermal instability at 25~MV/m for $RRR>200$ .

There have been many attempts to identify defects which were localized by temperature mapping. Examples of defects are drying spots, fibers from tissues, foreign material inclusions, weld splatter and cracks in the welds. There are two obvious and successful methods for reducing the danger of thermal instability: 
\begin{itemize}
\item
avoid defects by preparing and cleaning the
cavity surface with extreme care;
\item
increase the thermal conductivity of the superconductor.
\end{itemize}
Considerable progress has been achieved in both aspects over the last ten years. 

Field emission of electrons from sharp tips is the most severe limitation in high-gradient superconducting cavities. In field-emission loaded cavities the quality factor drops exponentially above a certain threshold, and X-rays are observed. 
The field emission current density is given by the Fowler-Nordheim equation \cite{Fowler1928}:

\begin{equation}\label{e:Fowler Nordheim}
j_{FE} = c_1E_{loc}^{2.5} \exp\left(-\frac{c_2}{\beta E_{loc}}\right)
\end{equation}
where $E_{loc}$ is the local electric field, $\beta$ a so-called field enhancement factor and $c_1$, $c_2$ are constants. There is experimental evidence that small particles on the cavity surface (e.g. dust) act as field emitters. Therefore perfect cleaning, for example by high-pressure water rinsing, is the most effective remedy against field emission. By applying this technique it has been possible to raise the threshold for field emission in multicell cavities from about 10 MV/m to more than 20 MV/m in the past few years.

The topics of thermal instability and field emission are discussed at much greater detail in the book by Padamsee, Knobloch and Hayes \cite{Padamsee98}. 

\section{Design of the TESLA Cavities}\label{design}

\subsection{Overview}

The TTF cavity is a 9-cell standing wave structure of about 1~m length whose lowest TM mode resonates at 1300 MHz. A photograph is shown in Fig.~\ref{cav_bild}. The cavity is made from solid niobium and is cooled by superfluid helium at 2 K.
\begin{figure}
\begin{center}
\caption{\label{cav_bild}Superconducting 1.3 GHz 9-cell cavity for the TESLA Test Facility.}
\end{center}
\end{figure}

Each 9-cell cavity is equipped with its own titanium helium tank, a tuning system driven by a stepping motor, a coaxial rf power coupler capable of transmitting more than 200 kW, a pickup probe and two higher order mode couplers. To reduce the cost for cryogenic installations, eight cavities and a superconducting quadrupole are mounted in a common vacuum vessel and constitute the so-called cryomodule of the TTF linac, shown in Fig.~\ref{module}. Within the module the cavity beam pipes are joined by stainless steel bellows and flanges with metallic gaskets. The cavities are attached to a rigid 300 mm diameter helium supply tube which provides positional accuracy of the cavity axes of better than 0.5 mm. Invar rods ensure that the distance between adjacent cavities remains constant during cooldown. Radiation shields at 5~K and 60~K together with 30 layers of superinsulation limit the static heat load on the 2~K level to less than  3~W for the 12~m long module.
\begin{figure}
\begin{center}
\caption{\label{module}Cryogenic module of the TESLA Test Facility linac comprising eight 9-cell cavities and a superconducting quadrupole.}
\end{center}
\end{figure}

\subsection{Layout of the TESLA cavities}

\subsubsection{Choice of frequency}

The losses in a microwave cavity are proportional to the product of conductor area and surface resistance. For a given length of a multicell resonator, the area scales with $1/f$ while the surface resistance of a superconducting cavity scales with $f^2$ for $R_{\rm BCS} \gg R_{\rm res}$ and is  independent of $f$ for $R_{\rm BCS}\ll R_{\rm res}$. At an operating temperature $T=2$ K the BCS term dominates above 3 GHz and hence the losses grow linearly with frequency whereas for frequencies below 300~MHz the residual resistance dominates and the losses grow with $1/f$. To minimize the dissipation in the cavity wall one should therefore select $f$ in the range 300 \,MHz to 3\,GHz.

Cavities in the 350 to 500~MHz regime are in use in electron-positron storage rings. Their large size is advantageous to suppress wake field effects and higher order mode losses. However, for a linac of several 10~km length the niobium
and cryostat costs for these bulky cavities would be prohibitive, hence a higher frequency has to be chosen. Considering material costs $f = 3$\,GHz  might
appear the optimum but there are compelling arguments for choosing about half this frequency. 
\begin{itemize}
\item
The wake fields losses scale with the second to third power of the frequency ($W_{\parallel}\propto f^2$, $W_{\perp}\propto f^3$). Beam emittance growth and beam-induced cryogenic losses are therefore much higher at 3 GHz.

\item
The $f^2$ dependence of the BCS resistance sets an upper limit\footnote{See Fig. 11.22 in \cite{Padamsee98}.} of about 30\,MV/m at 3\,GHz, hence choosing this frequency would definitely preclude a possible upgrade of TESLA to 35--40\,MV/m \cite{Padamsee91}. 
\end{itemize}

The choice for 1.3\,GHz was motivated by the availability of high power klystrons.

\subsubsection{Cavity geometry}

\begin{figure}
\begin{center}
\epsfig{figure=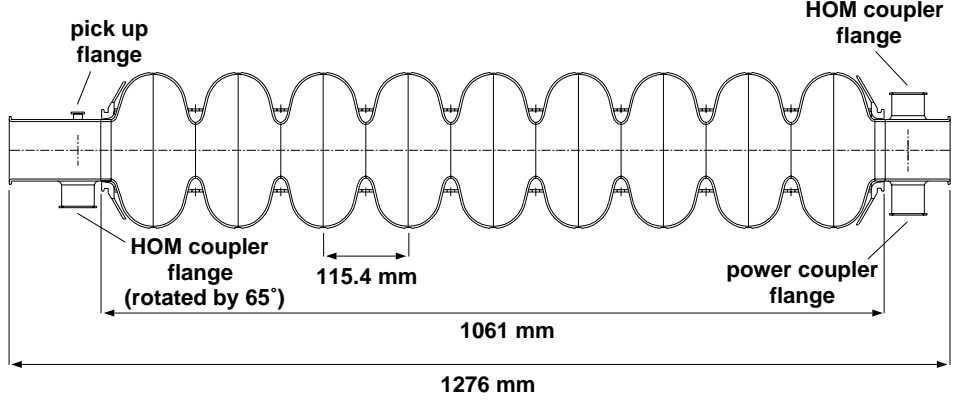,scale=0.5,angle=-90}
\caption { \label{FC1}Side view
of the 9-cell TTF cavity with the ports for the main power coupler and two higher-order mode (HOM) couplers.}
\end{center}
\end{figure}

A multicell resonator is advantageous for maximizing the active acceleration length in a linac of a given size. With increasing number of cells per cavity, however, difficulties arise from trapped modes, uneven field distribution in the cells and too high power requirements on the input coupler. Extrapolating from the experience with 4-cell and 5-cell cavities a 9-cell structure appeared manageable. 
A side view of the TTF cavity with the beam tube sections and the coupler ports is given in Fig.~\ref{FC1}.

The design of the cell shape was guided by the following considerations:
\begin{itemize}
\item
a spherical contour near the equator with low sensitivity for multipacting,
\item
minimization of electric and magnetic fields at the cavity wall
to reduce the danger of field emission and thermal breakdown,
\item
a large iris radius to reduce wake field effects.
\end{itemize}

The shape of the cell was optimized using the code URMEL \cite{urm}. The resonator is operated in the $\pi$ mode with $180^\circ$ phase difference between adjacent cells. The longitudinal dimensions are determined by the condition that the electric field has to be inverted in the time a relativistic particle needs to travel  from one cell to the next. The separation between two irises is therefore  $ c/(2f)$. The iris radius $R_{\rm{iris}}$ influences the cell-to-cell coupling\footnote{The coupling coefficient is related to the frequencies of the coupled modes in the 9-cell resonator by the formula \mbox{$f_n = f_0/ \sqrt{1+2\, k_{\rm cell} \cos(n\pi/9)}$} where $f_0$ is the resonant frequency of a single cell and $1 \le n \le 9$.} $k_{\rm cell}$, the excitation of higher order modes and other important cavity parameters, such  as the ratio of the peak electric (magnetic) field at the cavity wall to the accelerating field and the ratio $(R/Q)$ of shunt impedance  to quality factor. For the TESLA Test Facility cavities $R_{\rm{iris}}=35$~mm was chosen, leading to $k_{\rm cell}$~=~1.87\,\% and $E_{\rm peak}/E_{\rm acc}$~=~2. 
The most important parameters are listed in Table~\ref{t:tescav}.

\begin{table}
\begin{center}
\begin{minipage}{\linewidth}
\caption{\label{t:tescav} TTF cavity design parameters.\protect\footnote{Following common usage in ac circuits and the convention adopted in the Handbook of Accelerator Physics and Engineering \cite{Chao99}, page 523, we define the shunt impedance by the relation $R=V^2/(2P)$, where $P$ is the dissipated power and $V$ the peak voltage in the equivalent parallel LCR circuit. Note that another definition is common, which has also been used in the TESLA Conceptual Design Report: $R=V^2/P$, leading to a factor of 2 larger shunt impedance.}}
\begin{tabular}{|l|c|}
\hline
type of accelerating structure & standing wave \\ \hline
accelerating mode & TM$_{010}\; $, $\pi$ mode \\ \hline
fundamental frequency & 1300 MHz \\ \hline
design gradient $E_{acc}$ & 25 MV/m \\ \hline 
quality factor $Q_0$ & $>$ 5 $\cdot 10^9$ \\ \hline
active length $L$ & 1.038 m \\ \hline
number of cells & 9 \\ \hline
cell-to-cell coupling & 1.87 \% \\ \hline
iris diameter & 70 mm \\ \hline
geometry factor & 270 $\Omega$ \\ \hline
$R/Q$ & 518 $\Omega$ \\ \hline
$E_{\rm peak}/E_{\rm acc}$ & 2.0 \\ \hline
$B_{\rm peak}/E_{\rm acc}$  & 4.26 mT/(MV/m) \\ \hline
tuning range & $\pm$ 300 kHz \\ \hline
$\Delta f/\Delta L$ & 315 kHz/mm \\ \hline
Lorentz force detuning at 25 MV/m & $\approx$ 600 Hz \\ \hline
$Q_{\rm ext}$ of input coupler & 3 $\cdot 10^6$ \\ \hline
cavity bandwidth at $Q_{\rm ext} = 3 \cdot 10^6$ & 430 Hz \\ \hline
RF pulse duration & 1330 $\mu$s \\ \hline
repetition rate & 5 Hz \\ \hline
fill time & 530 $\mu$s \\ \hline
beam acceleration time & 800 $\mu$s \\ \hline
RF power peak/average & 208 kW/1.4 kW \\ \hline
number of HOM couplers & 2 \\ \hline
cavity longitudinal loss factor k$_{\parallel}$ for $\sigma_z$ = 0.7 mm &
10.2 V/pC \\ \hline
cavity transversal loss factor k$_{\perp}$ for $\sigma_z$ = 0.7 mm & 
15.1 V/pC/m \\ \hline
parasitic modes with the highest impedance : \hspace{5mm} type & 
TM$_{011}$ \\ 
\makebox[40mm]{} $\pi$/9 \hspace{5mm} ($R/Q$)/ \hspace{5mm}frequency 
& 80 $\Omega$/2454 MHz \\ 
\makebox[40mm]{} 2$\pi$/9 \hspace{3mm} ($R/Q$)/ \hspace{5mm} frequency 
& 67 $\Omega$/2443 MHz \\ \hline
bellows longitudinal loss factor k$_{\parallel}$ for $\sigma_z$ = 0.7 mm &
1.54 V/pC \\ \hline
bellows transversal loss factor k$_{\perp}$ for $\sigma_z$ = 0.7 mm &
1.97 V/pC/m \\ \hline 
\end{tabular}
\end{minipage}
\end{center}
\end{table}

\begin{figure}
\begin{center}
\epsfig{figure=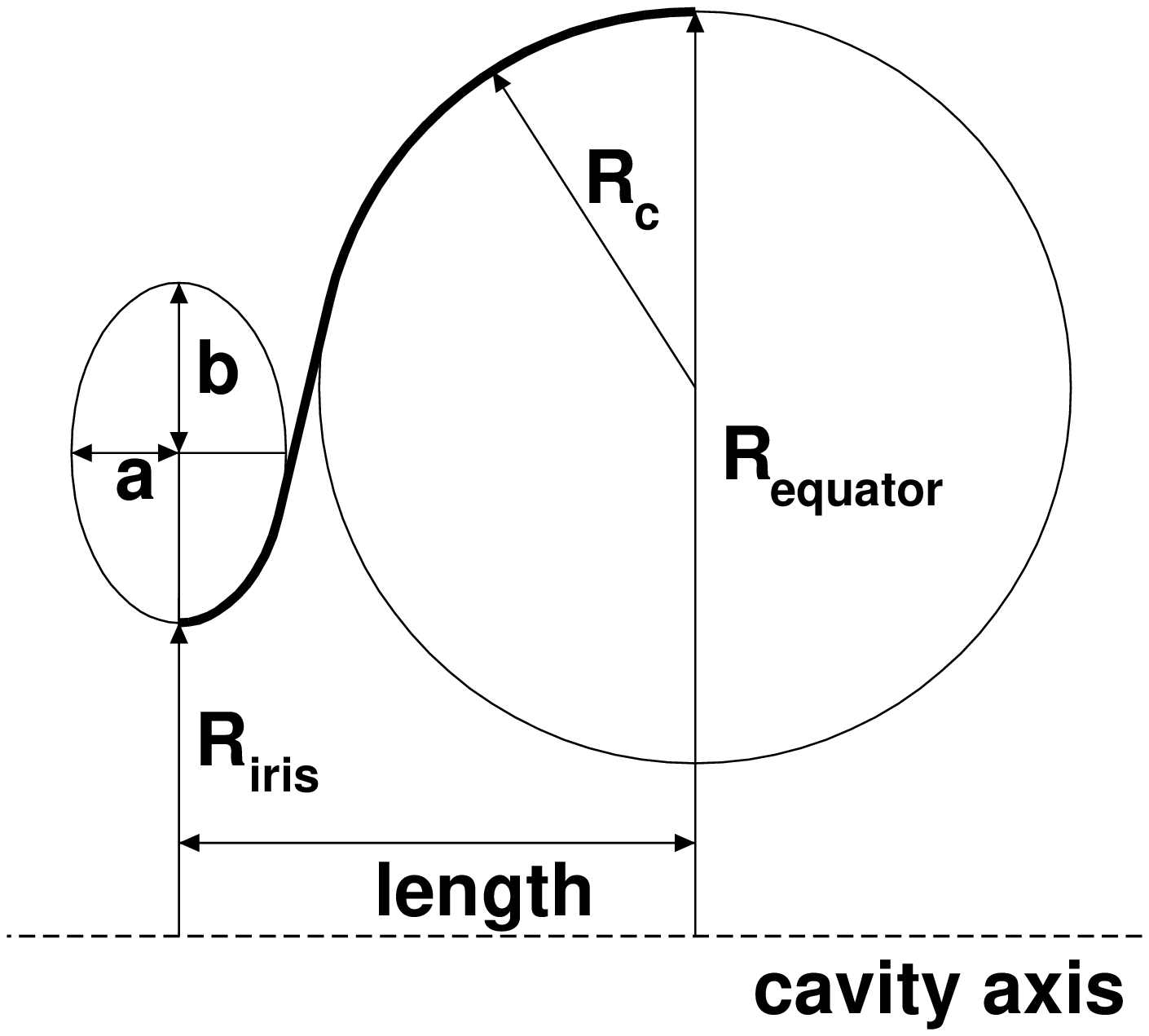,height=9cm,angle=-90}
\caption{\label{FC2}Contour of a half cell.}
\end{center}
\end{figure}

The contour of  a half-cell is shown in Fig.~\ref{FC2}. It is composed of a circular arc around the equator region and an elliptical section near the iris. The dimensions are listed in Table~\ref{t:cavshape}. The half-cells at the end of the 9-cell resonator need a slightly different shape to ensure equal field amplitudes in all 9 cells. In addition there is a slight asymmetry between left and right end cell which prevents trapping of higher-order modes (see Sect. 3.5).

\begin{table}[hbt]
\begin{center}
\caption{\label{t:cavshape}Half-cell shape parameters 
 (all dimensions in mm).}
\begin{tabular}{|l|l|l|l|}
\hline
cavity
shape parameter & midcup & endcup 1 & endcup 2 \\ 
\hline\hline
equator
radius $R_{\rm equat.}$ & 103.3 & 103.3 & 103.3 \\ \hline
iris radius $R_{\rm iris}$ &  35 &
39 &   39 \\ \hline
radius $R_{\rm arc}$ of circular arc&  42.0 &  40.3 &   42 \\
\hline
horizontal half axis $a$ &  12 &  10 &    9 \\ \hline
vertical half
axis $b$ &  19 &  13.5 &  12.8 \\ \hline
length $l$ &  57.7 &  56.0 &  57.0
\\ \hline
\end{tabular}
\end{center}
\end{table}

\subsubsection{ Lorentz-force detuning and cavity stiffening}

\begin{figure}
\begin{center}
\epsfig{figure=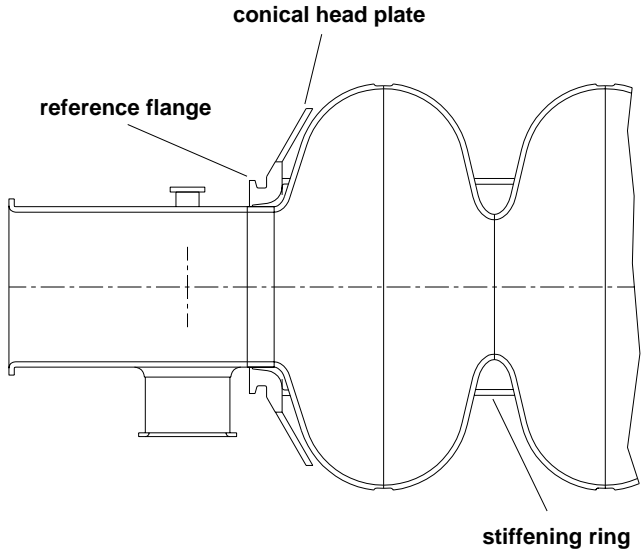,scale=0.5,angle=-90}
\caption{\label{FC7}End section of a cavity with stiffening ring, conical 
head plate for welding into the helium tank and reference flange for 
alignment.}
\end{center}
\end{figure}

The electromagnetic field exerts a Lorentz force on the currents induced in a thin surface layer. The resulting pressure acting on the cavity wall
\begin{equation}
p=\frac{1}{4}(\mu_{0}H^{2}-\varepsilon_{0}E^{2})
\end{equation}
leads to a deformation of the cells in the $\mu$m range and a change $\Delta V$ of their volume. The consequence is a frequency shift according to Slater's rule
\begin{equation}\label{e:fshift}
\frac{\Delta f}{f_{0}}=\frac{1}{4W}
\int_{\Delta V} (\varepsilon_{0}E^{2}-\mu_{0}H^{2})dV\;.
\end{equation}
Here
\begin{equation}
W=\frac{1}{4}\int_{V} (\varepsilon_{0}E^ {2}+\mu_{0}H^{2})dV
\end{equation}
is the stored energy and $f_{0}$ the resonant frequency of the unperturbed cavity. The computed frequency shift at 25\,MV/m amounts to 900\,Hz for an unstiffened cavity of 2.5\,mm wall thickness. The
bandwidth of the cavity equipped with the main power coupler
($Q_{\rm ext}=3\cdot 10^6$)  is about 430\,Hz, hence a reinforcement of the cavity is needed. Niobium stiffening rings are welded in
between adjacent cells as shown in Fig.~\ref{FC7}. They reduce the
frequency shift to about 500\,Hz for a 1.3\,ms long rf 
pulse\footnote{Part of this shift is due to an elastic deformation of the tuning mechanism.}, see Fig.~\ref{F6.3}. 

The deformation of the stiffened cell is negligible near the iris where the electric field is large, but remains  nearly the same as in  the unstiffened cell near the equator where the magnetic field dominates. The deformation in this region can only be reduced by increasing the wall thickness.

\subsubsection{Magnetic Shielding\label{s:magneticshielding}}

\begin{figure}
\begin{center}
\epsfig{figure=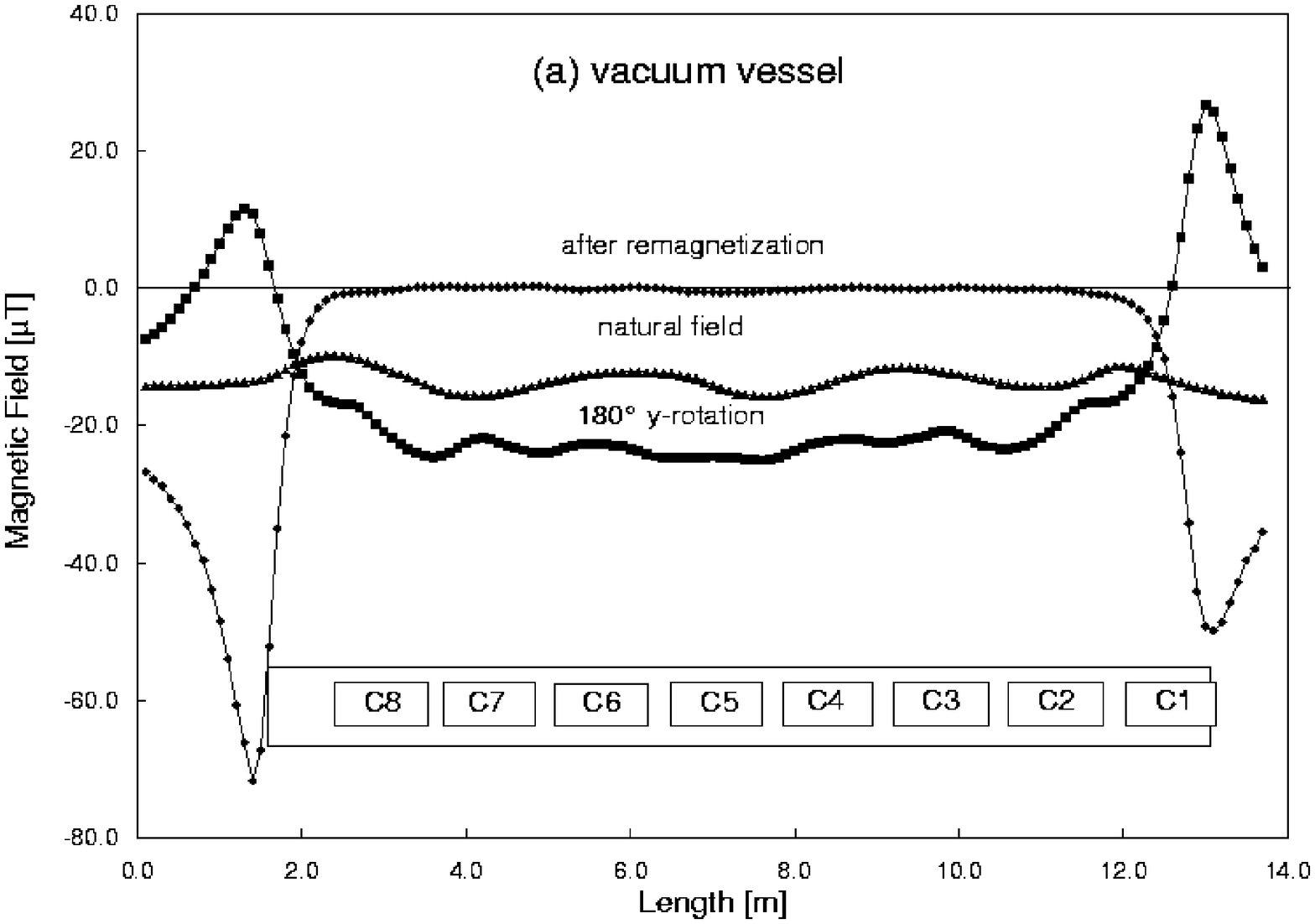,width=9cm,height=14.5cm,angle=-90}

\epsfig{figure=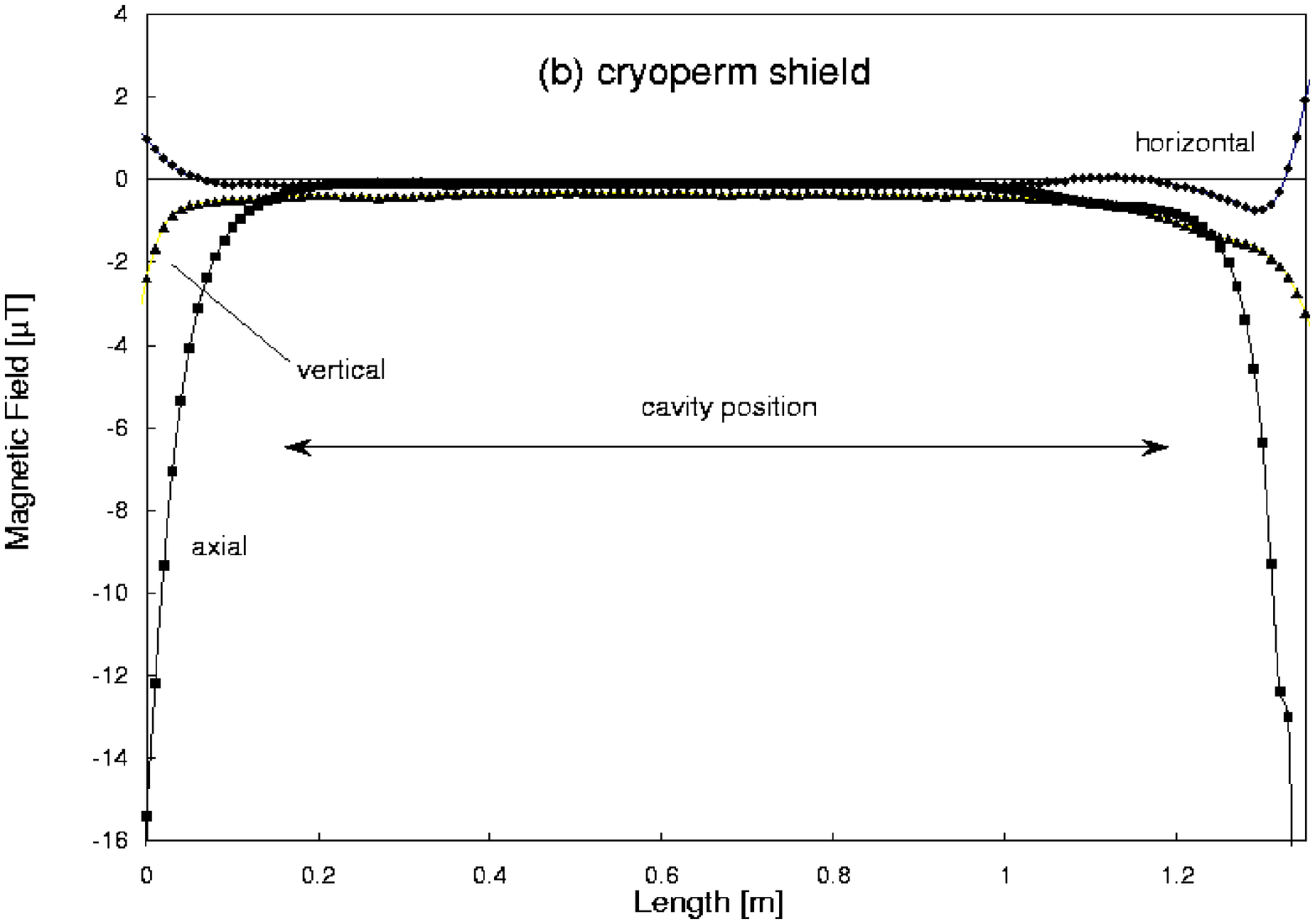,width=9cm,height=14.5cm,angle=-90}

\caption{\label{FC11}Shielding of the Earth's magnetic field. 
(a) Shielding of axial component by the steel vacuum vessel of the cryomodule. Shown is also the arrangement of the cavity string in the vessel. 
(b) Shielding of the axial, horizontal and vertical field components by the cryoperm cylinder surrounding the cavity (measured without vacuum vessel).}
\end{center}
\end{figure}

As shown in Sect.~\ref{shielding} the ambient magnetic field must be shielded to a level of about a $\mu$T to reduce the magnetic surface resistance to a few n$\Omega$. This is accomplished with a two-stage passive shielding, provided by the conventional steel vacuum vessel of the cryomodule and a high-permeability cylinder around each cavity. To remove the remanence from the steel vessel the usual demagnetization technique is applied. 
The resulting attenuation of the ambient field is found to be 
better than expected from a cylinder without any remanence. The  
explanation is that the procedure does not really demagnetize the steel but rather {\it remagnetizes} it in such a way that  the axial component of the ambient field is counteracted. This interpretation (see also ref. \cite{albach_voss}) becomes obvious if the cylinder is turned by 180$^\circ$: 
in that case the axial field measured inside the steel cylinder is almost twice as large as the ambient longitudinal field component, see Fig.~\ref{FC11}a. 

The shielding cylinders of the cavities are made from Cryoperm\footnote{Cryoperm is made by Vacuumschmelze Hanau, Germany.} 
which retains a high permeability of more than 10000 when 
cooled to liquid helium temperature. Figure~\ref{FC11}b  shows the measured horizontal, vertical and axial components inside a cryoperm shield at room temperature, which was exposed to the Earth's field. The combined action of
remagnetized vacuum vessel and cryoperm shield is more than
adequate to reduce the ambient field to the level of some $\mu$T. An
exception are the end cells of the first and the last cavity near the end of
the cryomodule where the vessel is not effective in attenuating
longitudinal fields. Here an active field compensation by means of Helmholtz coils could reduce the fringe field at the last cavity to a harmless level. 

\subsection{Helium vessel and tuning system\label{ss:heliumvess}}

The helium tank contains the superfluid helium needed for cooling and serves at the same time as a mechanical support of the cavity and as a part of the tuning mechanism. The tank is made from titanium whose differential thermal contraction relative to niobium is 20 times smaller  than for stainless steel.  Cooldown produces a stress of only 3 MPa  in a cavity that was stress-free at room
temperature. Titanium has the additional advantage that it can be 
directly electron-beam welded to niobium while stainless steel-niobium 
joints would require an intermediate metal layer.

The assembly of cavity and helium tank proceeds
in the following sequence:  a titanium bellows is electron-beam (EB) welded to the conical Nb head plate at one side of the cavity, a titanium ring is EB welded to the conical Nb head plate at other side (see Fig. \ref{FC7}).  The cavity is then  inserted into the tank and the bellows as well as the titanium ring are TIG welded to the Ti vessel.
 
The tuning system consists of a stepping motor with a gear box and a double lever
arm. The moving parts operate at 2\,K in vacuum. The tuning
range is about $\pm$1\,mm, corresponding to a frequency range of 
$\pm$300\,kHz. The resolution is 1\,Hz. The tuning system is adjusted in such
a way that after cooldown the cavity is always under compressive force to
avoid a backlash if the force changes from pushing to
pulling.

\subsection{Main Power Coupler}

\subsubsection*{Design requirements}
A critical component of a superconducting cavity is the power input coupler. For TTF several coaxial couplers have been developed  \cite{Moeller99}, consisting
of a ``cold part'' which is mounted on the cavity in the clean room and closed by a ceramic window, and a ``warm part'' which is assembled after
installation of the cavity in the cryomodule. The warm section contains the transition from waveguide to coaxial line. This part is evacuated and
sealed against the air-filled wave guide by a second ceramic window. The elaborate two-window solution was chosen to get optimum protection of the cavity against contamination during mounting in the cryomodule and against window fracture during linac operation. 

The couplers must allow for some longitudinal motion\footnote{The motion of the coupler ports is up to 15~mm in the first cryomodules but has been reduced to about 1~mm in the most recent cryostat design by fixing the distance between neighboring cavities with invar rods.} inside the 12\,m long cryomodule when the cavities are cooled down from room temperature to 2\,K. For this reason bellows in the inner and outer conductors of the coaxial line are needed. Since the coupler connects the room-temperature waveguide with the 2\,K cavity, a compromise must be found between a low thermal conductivity and a high electrical conductivity. This is
achieved by several thermal intercepts and  
by using stainless steel pipes or bellows with a thin copper plating (10--20 $\mu$m) at the radio frequency surface. The design heat loads of 6\,W at 70\,K, 0.5\,W at 4\,K and 0.06\,W at 2\,K have been undercut in practice. 

\subsubsection*{Electrical properties}

An instantaneous power of 210\,kW has to be transmitted to
provide a gradient of 25\,MV/m for an 800\,$\mu$s long beam pulse of 
8~mA. The filling 
time of the cavity amounts to 530\,$\mu$s and the decay time, after the beam pulse is over, to an additional 500~$\mu$s. At the beginning of the 
filling, most of the rf wave is reflected leading to voltage enhancements by a factor of 2. The external quality factor of the coupler is $Q_{ext} = 3\cdot 10^6$ at 25 MV/m. By moving the inner conductor of the coaxial line, $Q_{ext}$ can be varied in the range $ 1\cdot 10^6$ -- $9\cdot 10^6$ to allow not only for different beam loading conditions but also to facilitate an in-situ high power processing of the cavities. This feature has proved extremely useful on several occasions to eliminate field emitters that entered the cavities at the last assembly stage.

\subsubsection*{Input coupler A \label{sss:fermicoup}}

The coupler version A is shown in Fig.~\ref{FC8}. It has a conical 
ceramic window at 70~K and a commercial planar waveguide window at room temperature. 
\begin{figure}
\begin{center}
\caption{\label{FC8}Simplified view of the power input coupler version A.}
\end{center}
\end{figure} 

A conical shape was chosen for the cold ceramic window to 
obtain broad-band impedance matching. The Hewlett-Packard High 
Frequency Structure Simulator program HFSS was used to
model the window and to optimize the shape of the tapered inner conductor.
 The reflected power is below 1\,\%. The ceramic window is made from Al$_2$O$_3$ with a purity of 99.5\%.  OFHC copper rings are brazed to the ceramic using  Au/Cu (35\%/65\%) braze alloy. The inner conductors on each side of the ceramic are electron-beam welded, the outer conductors are TIG welded. The ceramic is coated on both sides with a 10~nm titanium nitride layer to reduce multipacting.

The waveguide-to-coaxial transition is  realized using a cylindrical knob as the 
impedance-transforming device and a planar waveguide window. Matching posts are required on the air side of the window for impedance matching at 1.3~GHz.

\subsubsection*{\label{sss:desycoup}Input couplers B, C}

Coupler version B uses also a planar wave guide window and a door-knob transition from the wave guide to the coaxial line, but a cylindrical ceramic window at 70~K without direct view of the beam. Owing to a shortage in commercial wave guide windows a third type, C, was developed using a cylindrical window also at the wave guide - coaxial transition. It features a 60~mm diameter coaxial line with reduced sensitivity to multipacting and the possibility to apply a dc potential to the center conductor. In case of the LEP couplers \cite{Tue} a dc bias has proved very beneficial in suppressing multipacting. Similar observations were made at DESY. All couplers needed some conditioning but have then performed according to specification. 

\subsection{Higher order modes}

\begin{figure}
\begin{center}
\epsfig{figure=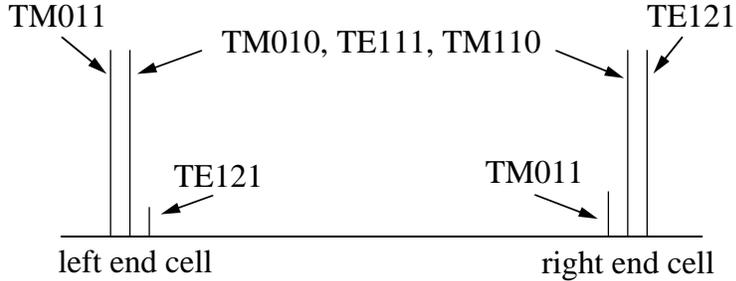,width=10cm}
\caption{\label{FC9}Effect of asymmetric end cell shaping on various 
modes. The main accelerating mode TM$_{010}$ and the higher modes TE$_{111}$ and 
TM$_{110}$ are not affected while TM$_{011}$ is enhanced in the left 
end cell, TE$_{121}$ in the right end cell. Using HOM couplers at both ends, all higher order modes can be extracted.}
\end{center}
\end{figure}

The intense electron bunches excite eigenmodes of higher frequency in the resonator which must be damped to avoid multibunch instabilities and beam breakup. This is accomplished by extracting the stored energy via higher-order mode (HOM) couplers mounted on the beam pipe sections of the nine-cell resonator. A problem arises from ``trapped modes'' which are
concentrated in the center cells and have a low field amplitude in the end
cells. An example is the TE$_{121}$ mode. By an asymmetric
shaping of the end half cells one can enhance the field amplitude of the
TE$_{121}$ mode  in one end cell while preserving the ``field flatness'' of
the fundamental mode and also the good coupling of the HOM couplers to the untrapped modes TE$_{111}$, TM$_{110}$ and TM$_{011}$.
The effects of asymmetric end cell tuning are sketched in Fig.~\ref{FC9}.

The two polarization states of dipole modes would in principle require two orthogonal HOM couplers at each side of the cavity. In a string of cavities, however, this complexity can be avoided since the task of the ``orthogonal'' HOM coupler can be taken over by the HOM coupler of the neighboring cavity. The viability of this idea was verified in measurements.

\subsubsection*{HOM coupler design}

The  HOM couplers are mounted at both ends of the cavity with a nearly perpendicular orientation\footnote{The angle between the two HOM couplers is not 
90$^{\circ}$ but 115$^{\circ}$ to provide also damping of quadrupole 
modes.} to ensure damping of dipole modes of either
polarization. A 1.3~GHz notch filter is incorporated to prevent energy extraction from the accelerating mode. Two types of HOM couplers
have been developed and tested, one mounted on a flange, the other 
welded to the cavity.

\begin{figure}
\begin{center}
\epsfig{figure=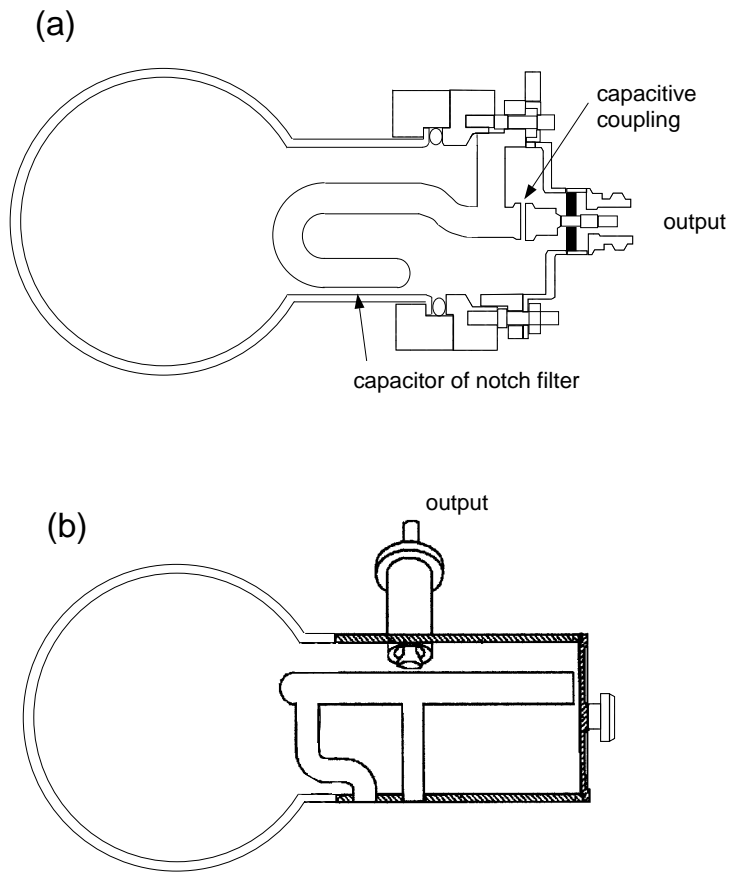,height=13cm}
\caption{\label{FC10}The higher-order-mode couplers: (a) demountable 
HOM coupler, (b) welded HOM coupler.}
\end{center}
\end{figure}

The demountable HOM coupler is shown in Fig.~\ref{FC10}a.
An antenna loop couples mainly to the magnetic field for TE modes and to the electric field for TM modes. The pickup antenna is capacitively coupled to an external load. The 1.3~GHz notch filter is formed by the inductance of the loop and the capacity at the 1.9~mm wide gap between loop and wall. A niobium bellows permits tuning of the filter without opening the cavity vacuum. The antenna is thermally connected to the 2\,K helium bath. In a cw (continuous wave) test at an accelerating field of 21\,MV/m the antenna reached a maximum temperature of 4\,K, which is totally uncritical.

The welded version of the HOM coupler is shown in Fig.~\ref{FC10}b. It resembles the couplers used in the 500\,MHz HERA cavities which have been operating for several years without quenches. The good cooling of the superconducting inner conductor by two stubs makes the design insensitive to $\gamma$ radiation and electron bombardment.

Both HOM couplers permit tuning of the fundamental mode rejection filter when mounted on the cavity. It is possible to achieve a $Q_{ext}$ of more than 
$ 10^{11}$ thereby limiting power extraction to less than 50~mW at 25~MV/m.  

\section{Cavity Fabrication and Preparation \label{fabrication}}

\subsection{Cavity fabrication}

\subsubsection{Niobium properties}
The 9-cell resonators are made from 2.8 mm thick sheet niobium by deep drawing of half cells, followed by trimming and electron beam welding. Niobium of high purity is needed. Tantalum with a typical concentration of 500 ppm is the most important metallic impurity. Among the interstitially dissolved impurities oxygen is dominant due to the high affinity of Nb for O$_2$ above 200$^\circ$C. Interstitial atoms act as scattering centers for the unpaired electrons and  reduce the $RRR$ and the thermal conductivity, see Sect. 2.1. The niobium ingot is highly purified by several remelting steps in a high vacuum electron beam furnace. This procedure reduces the interstitial oxygen, nitrogen and carbon contamination to a few ppm. The niobium specification for the TTF cavities is listed in Table \ref{Nb_spec}.

\begin{table}
\begin{center}
\caption{\label{Nb_spec}Technical specification for niobium used in TTF cavities}
\begin{tabular}{|l|l|l|l||l|l|}
\hline
\multicolumn{4}{|c|}{Impurity content in ppm (wt)} & \multicolumn{2}{|c|}{Mechanical Properties} \\ \hline
Ta & $\le 500$ & H & $\le 2$ & Residual resistivity ratio $RRR$ & $\geq 300$ \\ \hline
W & $\le 70$ & N & $\le 10$ & grain size & $\approx 50$ $\mu$m \\ \hline
Ti & $\le50$ & O & $\le 10$ & yield strength & $> 50$ MPa \\ \hline
Fe & $\le30$ & C & $\le 10$ & tensile strength & $> 100$ MPa \\ \hline
Mo & $\le50$ & & & elongation at break & 30 \% \\ \hline
Ni & $\le30$ & & & Vickers hardness HV 10 & $\le50$ \\ \hline
\end{tabular}
\end{center}
\end{table}			

After forging and sheet rolling, the 2.8 mm thick Nb sheets are degreased, a 
5~$\mu$m surface layer is removed by etching and then the sheets are annealed for 1--2 hours at 700--800$^{\circ}$C in a vacuum oven at a pressure of $10^{-5}$ -- $10^{-6}$ mbar to achieve full recrystallization and a uniform grain size of about 50 $\mu$m.  

\subsubsection{Deep drawing and electron-beam welding}

Half-cells are produced by deep-drawing. The dies are usually made from a high yield strength aluminum alloy. To achieve the small curvature required at the iris an additional step of forming, e.g. coining, may be needed. The half-cells are machined at the iris and the equator.  At the iris the half cell is cut to the specified length (allowing for weld shrinkage) while at the equator an extra length of 1 mm is left  to retain the possibility of a precise length trimming of the dumb-bell after frequency measurement (see below). The accuracy of the shape is controlled by sandwiching the half-cell between two metal plates and measuring the resonance frequency. 
The half-cells are thoroughly cleaned by ultrasonic degreasing, 20 $\mu$m chemical etching and ultra-pure water rinsing. Two half-cells are then joined at the iris with an electron-beam (EB) weld to form a ``dumb-bell''. The EB welding is usually done from the inside to ensure a smooth weld seam at the location of the highest electric field in the resonator.  Since niobium is a strong getter material for oxygen it is important to carry out the EB welds in a sufficiently good vacuum. Tests have shown that $RRR=300$ niobium is not degraded by welding at a pressure of less than $5 \cdot 10^{-5}$mbar. 

The next step is the welding of the stiffening ring. Here the weld shrinkage may lead to a slight distortion of the cell shape which needs to be corrected. 
Afterwards, frequency measurements are made on the dumb-bells to determine the correct amount of trimming at the equators. After proper cleaning by a 30 $\mu$m etching the dumb-bells are visually inspected. Defects and foreign material imprints from previous fabrication steps are removed by grinding. After the inspection and proper cleaning (a few $\mu$m etching followed by ultra-clean water rinsing and clean room drying), eight dumb-bells and two beam-pipe sections with attached end half-cells are stacked in a precise fixture to carry out the equator welds which are done from the outside. The weld parameters are chosen to achieve full penetration. A reliable method for obtaining a smooth weld seam of a few mm width at the inner surface is to raster a slightly defocused beam in an elliptic pattern and to apply  50 \% of beam power during the first weld pass and 100 \% of beam power in the second pass. 

\subsection{Cavity treatment}

Experience has shown that a damage layer in the order of 100 $\mu$m has to be removed from the inner cavity surface to obtain good rf performance in the superconducting state. The standard method applied at DESY and many other laboratories is called Buffered Chemical Polishing (BCP), using an acid mixture of HF (48 \%), HNO$_3$ (65 \%) and H$_3$PO$_4$ (85 \%) in the ratio 1:1:2 (at CEBAF the ratio was 1:1:1). The preparation steps adopted at DESY for the industrially produced TTF cavities are as follows. A layer of 80 $\mu$m is removed by BCP from the inner surface, 30 $\mu$m from the outer surface\footnote{These numbers are determined by weighing the cavity before and after etching and represent therefore the average over the whole surface. Frequency measurements indicate that more material is etched away at the iris than at the equator.}. The cavities are rinsed with ultra-clean water and dried in a class 100 clean room. The next step is a two-hour annealing at 800$^{\circ}$C in an Ultra High Vacuum (UHV) oven which serves to remove dissolved hydrogen from the niobium and relieves mechanical stress in the material. In the initial phase of the TTF program many cavities were tested after this step, applying a 20 $\mu$m BCP and ultra-clean water rinsing before mounting in the cryostat and cooldown. 

Presently, the cavities are rinsed with clean water after the  800$^{\circ}$C treatment and then immediately transferred to another UHV oven in which they are heated to 1350--1400$^{\circ}$C. At this temperature, all dissolved gases diffuse out of the material and the $RRR$ increases by about a factor of 2 to values around 500. To capture the oxygen coming out of the niobium and to prevent oxidation by the residual gas in the oven (pressure $<10^{-7}$mbar) a thin titanium layer is evaporated on the inner and outer cavity surface, Ti being a stronger getter than Nb. The high-temperature treatment with Ti getter is often called post-purification. The titanium layer is removed afterwards by a 80 $\mu$m BCP of the inner surface.  
A BCP of about 30 $\mu$m is applied at the outer surface since the Kapitza resistance of  titanium-coated niobium immersed in superfluid helium is about a factor of 2 larger than that of pure niobium \cite{kapitza}.  
After final heat treatment and BCP the cavities are mechanically tuned to adjust the  resonance frequency to the design value and to obtain equal field amplitudes in all 9 cells. This is followed by a slight BCP, three steps of high-pressure water rinsing (100~ bar) and drying in a class 10 clean room. As a last step, the rf test is performed in a superfluid helium bath cryostat.

A severe drawback of the post-purification is the considerable grain growth accompanied with a softening of the niobium. Postpurified-treated cavities are quite vulnerable to plastic deformation and have to be handled with great care.   

\section{Results on Cavity Performance and Quality Control Measures \label{results}}

\subsection{Overview}
\begin{figure}
\begin{center}
\epsfig{figure=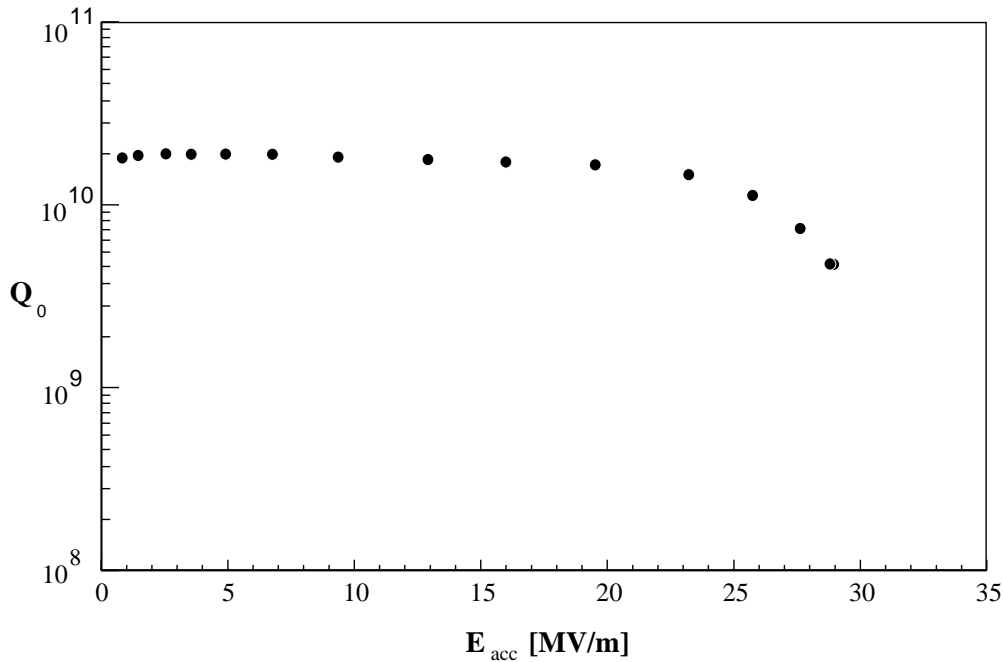,scale=0.55,angle=90}
\caption{\label{F4.1}
Excitation curve of the best TESLA 9-cell cavity measured up to date. The cavity was cooled by superfluid helium of 2 K.}
\end{center}
\end{figure}

Figure \ref{F4.1} shows the ``excitation curve'' of the best 9-cell resonator measured so far; plotted is the quality factor\footnote{The quality factor is defined as $Q_0=f/\Delta f$ where $f$ is the resonance frequency and $\Delta f$ the full width at half height of the resonance curve of the ``unloaded'' cavity.} $Q_0$ as a function of the accelerating electric field $E_{\rm acc}$. An almost constant and high value of $2 \cdot 10^{10}$ is observed up to 25 MV/m. 

\begin{figure}
\begin{center}
\epsfig{figure=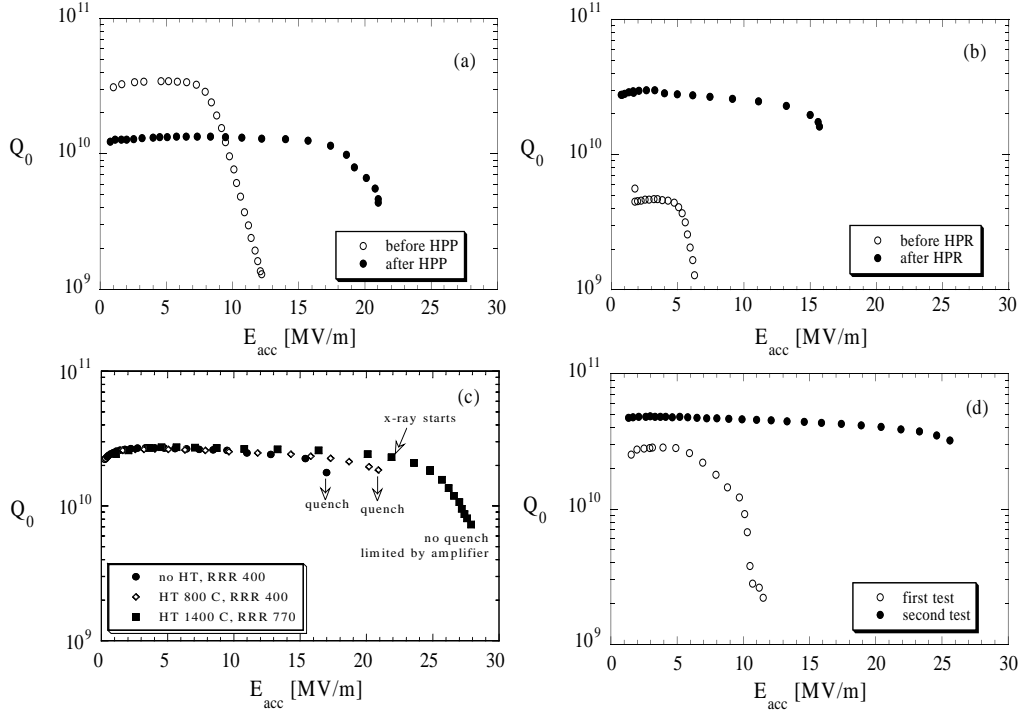,width=13.5cm,height=9.5cm}
\caption{\label{F4.2}Improvement in cavity performance due to various 
treatments: (a) high power processing, (b) high pressure water rinsing, 
(c) successive application of 800$^{\circ}$C and 1400$^{\circ}$C heat treatment, (d) removal of surface defects or titanium in grain boundaries by additional BCP.}
\end{center}
\end{figure}

The importance of various cavity treatment steps for arriving at such a good performance are illustrated in the next figure. A strong degradation is usually observed if a foreign particle is sticking on the cavity surface, leading either to field emission of electrons or to local overheating in the rf field. At Cornell University an {\it in situ} method for destroying field emitters was invented \cite{HPP}, called  ``high power processing'' (HPP), which in many cases can improve the high-field capability, see Fig. \ref{F4.2}a. Removal of field-emitting particles by high-pressure water rinsing, a technique developed at CERN \cite{HPR}, may dramatically improve the excitation curve (Fig. \ref{F4.2}b). The beneficial effect of a 1400$^{\circ}$C heat treatment, first tried out at Cornell \cite{HAT_corn} and Saclay \cite{HAT_Sac}, is seen in Fig. \ref{F4.2}c. Finally, an incomplete removal of the titanium surface layer in the BCP following the 1400$^\circ$C heat treatment may strongly limit the attainable gradient. Here additional BCP is of advantage (Fig. \ref{F4.2}d). 

\subsection{Results from the first series of TTF cavities}

After the successful test of two prototype nine-cell resonators a total of 27 cavities, equipped with main power and HOM coupler flanges, were ordered at four European companies. These cavities were foreseen for installation in the TTF linac with an expected gradient of at least 15 MV/m at $Q_0>3 \cdot 10^9$. However, in the specification given to the companies no guaranteed gradient was required. According to the test results obtained at TTF these resonators can be classified into four categories: 

\begin{description}
\item[\mbox{\normalfont(1)}] 16 cavities without any known material and fabrication defects, or with minor defects which could be repaired,
\item[\mbox{\normalfont(2)}]  3 cavities with serious material defects,
\item[\mbox{\normalfont(3)}]  6 cavities with imperfect equator welds, 
\item[\mbox{\normalfont(4)}]  2 cavities with serious fabrication defects (not fully penetrated electron beam welds or with holes burnt during welding; these were rejected).
\item One cavity has not yet been tested.
\end{description}

\begin{figure}
\begin{center}
\epsfig{figure=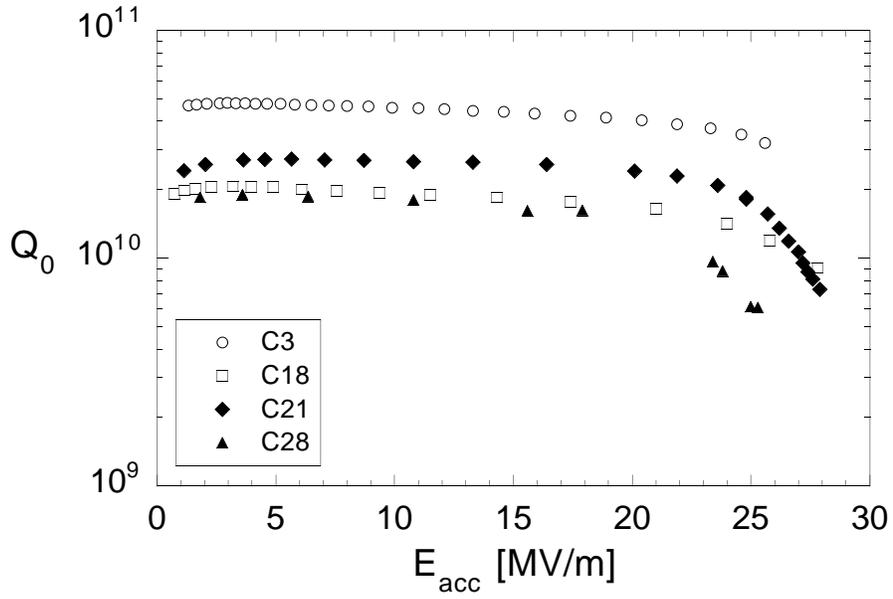,width=12cm}

(a)

\epsfig{figure=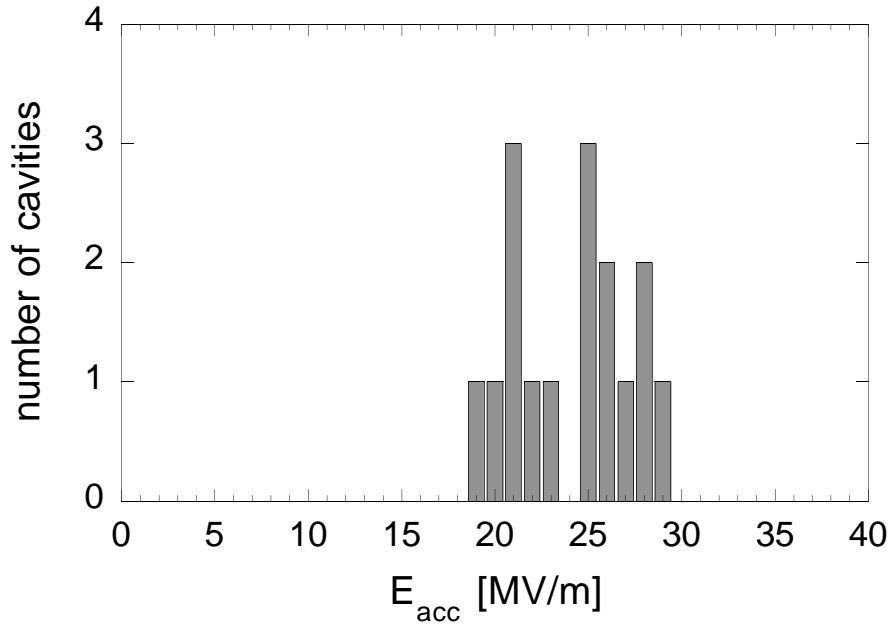,width=12cm}

(b)

\caption{\label{F4.3} (a) Excitation curves of the best 9-cell resonator of each of the four manufacturers. (b) Distribution of maximum gradients for the resonators of class 1, requiring a quality factor $Q_0 \ge 5\cdot 10^9$.}
\end{center}
\end{figure}

The test results for the cavities of class (1) in a vertical bath cryostat with superfluid helium cooling at 2 K are summarized in Fig. \ref{F4.3}. It is seen that the TTF design goal of 15 MV/m is clearly exceeded. Nine of the resonators fulfill even the more stringent specification of TESLA ($E_{\rm acc}\ge 25$ MV/m at $Q_0 \ge 5 \cdot 10^9$).

\begin{figure}
\begin{center}
\epsfig{figure=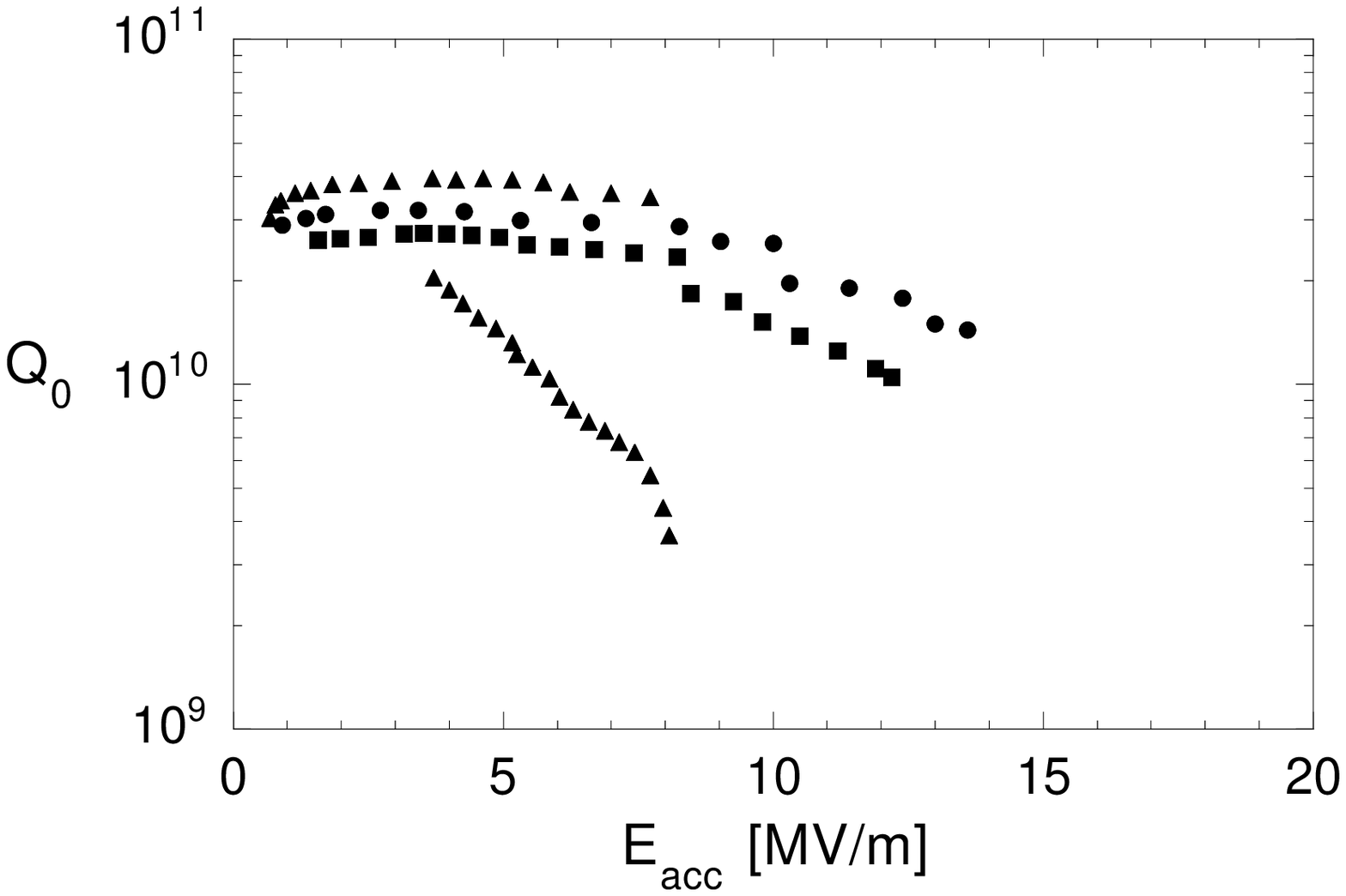,width=11cm,height=9cm}
\caption{\label{F4.4} Excitation curves of three cavities with serious material defects (class 2). Cavity C5 ($\blacktriangle$) exhibited a jump in quality factor.}
\end{center}
\end{figure}

The excitation curves of the class 2 cavities (Fig. \ref{F4.4}) are characterized by sudden drops in quality factor with increasing field and rather low maximum gradients.
Temperature mapping revealed spots of excessive heating at isolated spots which were far away from the EB welds. An example is shown in Fig. \ref{F4.5}a. The defective cell was cut from the resonator and subjected to further investigation \cite{Singer98}. An eddy-current scan, performed at the Bundesanstalt f\"ur Materialforschung (BAM) in Berlin, showed a pronounced signal at the defect location. With X-ray radiography, also carried out at BAM, a dark spot with a size of 0.2--0.3 mm was seen (Fig. \ref{F4.5}b) indicating an inclusion of foreign material with a higher nuclear charge than niobium. Neutron absorption measurements at the Forschungszentrum GKSS in Geesthacht gave no signal, indicating that the neutron absorption coefficient of the unknown contamination was similar to that of Nb. The identification of the foreign inclusion was finally accomplished using X-ray fluorescence (XAFS) at the Hamburger Synchrotronstrahlungslabor HASYLAB at DESY. Fluorescence was observed at photon energies corresponding to the characteristic X-ray lines of tantalum L$_1= 11.682$ keV, L$_2= 11.136$ keV and L$_3=9.881$ keV. 
\begin{figure}
\begin{center}
%\epsfig{figure=F12a_BWPh.eps,width=7.8cm}  ---> now in JPEG-Format
%$\quad$ \epsfig{figure=F12b_OJS.eps,width=5.1cm}  ---> now in JPEG-Format
$\qquad \qquad \qquad \qquad \quad$(a) $\qquad \qquad \qquad \qquad \qquad \qquad \qquad \qquad \quad$(b)
\caption{\label{F4.5} (a) Temperature map of cell 5 of cavity C6 showing excessive heating at a localized spot. (b) Positive print of an X-ray radiograph showing the ``hot spot'' as a dark point.}
\end{center}
\end{figure}
The SYRFA (synchrotron radiation fluorescence analysis) method features sufficient sensitivity to perform a scan of the tantalum contents in the niobium by looking at the lines Ta-K$_{\alpha 1}=57.532$ keV, Ta-K$_{\alpha 2}=56.277$ keV and Ta-K$_{\beta 1}=65.223$ keV. The average Ta content in the bulk Nb was about 200 ppm but rose to 2000 ppm in the spot region. The $RRR$ dropped correspondingly from 330 to about 60. 
 \begin{figure}
\begin{center}
\epsfig{figure=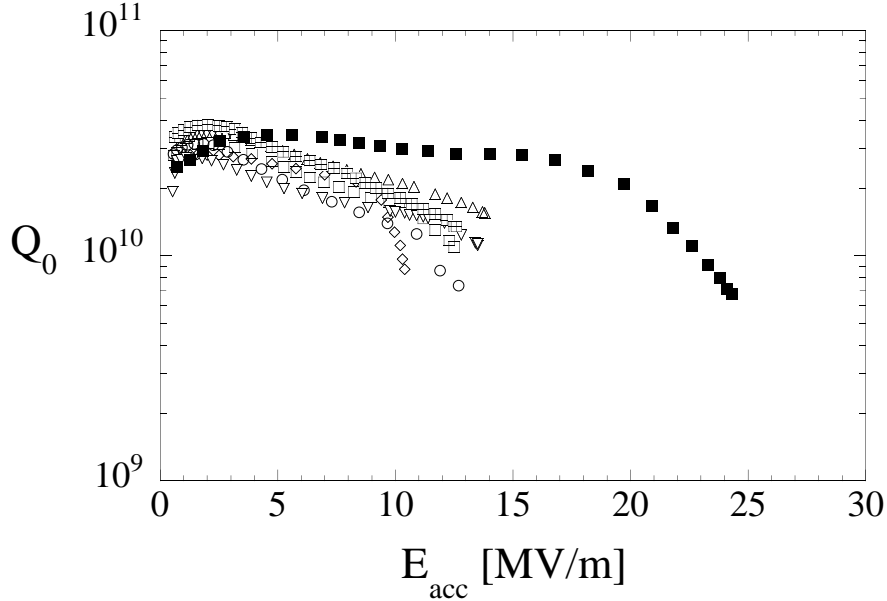,width=12cm}
\caption{\label{F4.7} Excitation curves of six cavities with imperfect equator welds (class 3). Also shown is a resonator ($\blacksquare$) made later by the same company, following stringent cleaning procedures at the equator welds.}
\end{center}
\end{figure}

\begin{figure}
\begin{center}
\epsfig{figure=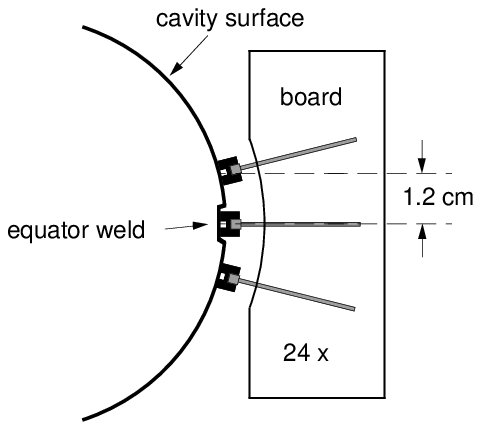,width=5cm}

(a)

\epsfig{figure=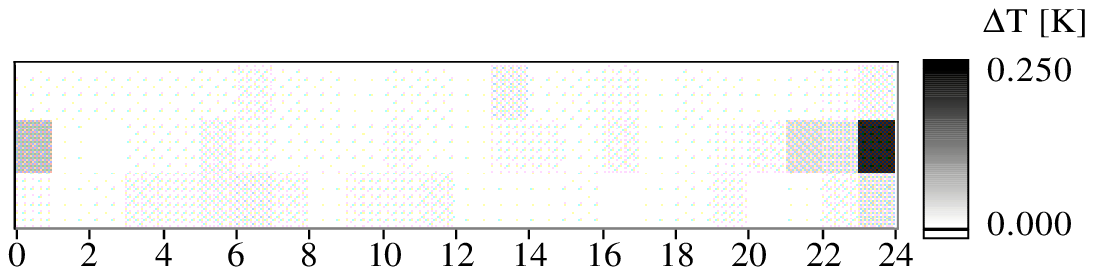,width=8cm}

(b)

\epsfig{figure=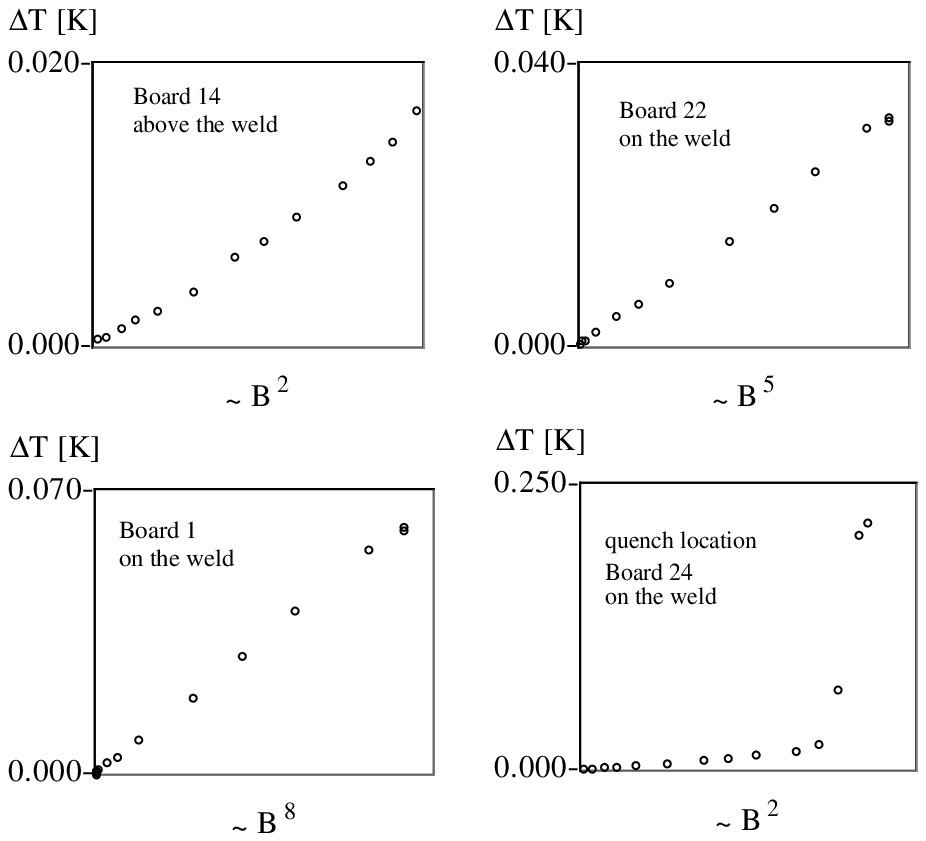,width=8cm}

(c)

\caption{\label{F4.8} (a) Location of temperature sensors to determine heating at the equator weld. \newline  (b) Temperature map of the equator region from cell 5 of cavity C9 just below the quench. \newline (c) Temperature rise at various locations as a function of $B^n$ with $n$ between 2 and 8.}
\end{center}
\end{figure}

The six cavities in class 3 were produced by one company and exhibited premature quenches at gradients of 10--14 MV/m and a slope in the $Q(E)$ curve (Fig.~\ref{F4.7}). Two of the resonators were investigated in greater detail \cite{Brinkmann98}. Temperature mapping revealed strong heating at several spots on the equator weld (Fig.~\ref{F4.8}b). The temperature rise as a function of the surface magnetic field is plotted in Fig.~\ref{F4.8}c for one sensor position above the weld and three positions on the weld. In the first case a growth proportional to $B^2$ is observed as expected for a constant surface resistance. On the weld, however, a much stronger rise is seen ranging from $B^5$ to $B^8$. This is clear evidence for a contamination of the weld seam.

Once the reason for the reduced performance of the cavities in class 3 had been identified a new 9-cell resonator was manufactured by the same company applying careful preparation steps of the weld region: a 2 $\mu$m chemical etching not more than 8 hours in advance of the EB welding, rinsing with ultrapure water and drying in a clean room. A rastered electron beam was used for welding with 50\% penetration in the first weld layer and 100\% in the second. The new cavity indeed showed excellent performance and achieved 24.5 MV/m, see Fig. \ref{F4.7}. The same applies for later cavities made by this company. 

The average gradient of the cavities without serious material or fabrication defects amounts to $20.1\pm6.2$ MV/m at $Q_0=5 \cdot 10^9$ where the error represents the rms of the distribution.

\subsection{Diagnostic methods and quality control} 

The deficiencies encountered in the first series production of TESLA cavities have initiated the development of diagnostic methods and quality control procedures. 

\subsubsection*{Electron microscopy}

Scanning electron microscopy with energy-dispersive X-ray analysis (EDX) is used to identify foreign elements on the surface. Only a depth of about 1~$\mu$m can be penetrated, so one has to remove layer by layer to determine the diffusion depth of titanium or other elements. Alternatively one can cut the material and scan the cut region. The titanium layer applied in the high temperature treatment has been found to extend to a depth of about 10~$\mu$m in the bulk niobium. The sensitivity of the EDX method is rather limited; a Ti fraction below 0.5~\% is undetectable. Auger electron spectroscopy offers higher sensitivity and using this method titanium migration at grain boundaries has been found to a depth of  50--100~$\mu$m. Hence this large thickness must be removed from the rf surface by BCP after post-purification with Ti getter. The detrimental effect of insufficient titanium removal has already been shown in Fig.~\ref{F4.2}d. The microscopic methods are restricted to small samples and cannot be used to study entire cavities. 

\subsubsection*{X-ray fluorescence}

The narrow-band X-ray beams at HASYLAB permit element 
identification via fluorescence analysis. In principle the existing apparatus allows the scanning of a whole niobium sheet such as used for
producing a half-cell, however the procedure would be far too time-consuming. 

\subsubsection*{Eddy-current scan}
\begin{figure}
\begin{center}
%\epsfig{file=F17.eps,width=7cm}  ---> now in GIF-Format
%$\quad$ \epsfig{file=FigEDA.eps,width=6.0cm}  ---> now in JPEG-Format
$\qquad \qquad \qquad \qquad \quad$(a) $\qquad \qquad \qquad \qquad \qquad \qquad \qquad \qquad \quad$(b)
\caption{\label{F4.10} (a) Schematic view of the xy eddy-current scanning system. (b) Photo of the new rotating scanning system.}
\end{center}
\end{figure}

A practical device for the quality control of all niobium sheets going into cavity production is a high-resolution eddy-current system developed by the Bundesanstalt f\"ur Materialforschung (BAM) in Berlin. The apparatus is shown in Fig. \ref{F4.10}. The frequency used is 100 kHz corresponding to a penetration depth of 0.5 mm in niobium at room temperature. The maximum scanning speed is 1 m/s. The scanning probe containing the inducing and receiving coils floats on an air pillow to avoid friction. The machined base plate contains holes for evacuating the space between this plate and the Nb sheet. The atmospheric pressure is sufficient to flatten the 265~x~265~mm$^2$ niobium sheets to within 0.1 mm which is important for a high sensitivity scan. The performance of the apparatus was tested with a Nb test sheet containing implanted tantalum deposits of 0.2 to 1 mm diameter. The scanned picture (Fig. \ref{F4.11}a) demonstrates that Ta clusters are clearly visible. Using this eddy-current apparatus the tantalum inclusion in cavity C6 was easily detectable. 

In the meantime an improved eddy-current scanning device has been designed and built at BAM which operates similar to a turn table and allows for much higher scanning speeds and better sensitivity since the accelerations of the probe head occuring in xy scans are avoided. A two frequency principle is applied in the new system. Scanning with high frequency (about 1 MHz) allows detection of surface irregularities while the low frequency test (about 150 kHz) is sensitive to bulk inclusions. The high and low frequency signals are picked up simultaneously. Very high sensitivity is achieved by signal subtraction.

\subsubsection*{Neutron activation analysis}

The eddy-current scan allows the detection of foreign materials in the niobium but is not suitable for identification. Neutron activation analysis permits a non-destructive determination of the contaminants provided they have radioactive isotopes with a sufficiently long half life. Experiments were carried out at the research reactor BER II of the Hahn Meitner Institut in Berlin. The niobium sheets are exposed to a thermal neutron flux of $10^9\,$cm$^{-2}$s$^{-1}$ for some 5 hours. The radioactive isotope $^{94}$Nb has a half life of 6.2 min while $^{182}$Ta has a much longer half life of 115 days. Two weeks after the irradiation the $^{94}$Nb activity has dropped to such a low level that tantalum fractions in the ppm range can be identified. Figure \ref{F4.11}b shows the implanted tantalum clusters in the specially prepared Nb plate with great clarity. Also the uniformly dissolved Ta is visible and the inferred concentration of 200 ppm is in agreement with the chemical analysis. The activation analysis is far too time consuming for series checks but can be quite useful in identifying special contaminations found with the eddy-current system. Ten Nb sheets from the regular production were investigated without showing any evidence for tantalum clusters.
\begin{figure}
\begin{center}
%\epsfig{file=Fig8Diagn.eps,width=6.5cm}  ---> now in JPEG-Format
%$\quad$ \epsfig{file=Fig10Diagn.eps,width=6.5cm}  ---> now in JPEG-Format

(a) $\qquad \qquad \qquad \qquad \qquad \qquad \qquad \quad$(b) 
\caption{\label{F4.11} (a) Eddy-current scan of a specially prepared Nb sheet with Ta implantations at 5 locations. (b) Neutron activation analysis of the same sheet.}
\end{center}
\end{figure} 

\subsection{Present status of TTF cavities}
\subsubsection{Improvements in cavity production}
For the second series 25 cavities have been ordered at four firms. The second production differed from the first one in three main aspects.

\subsubsection*{(1) Stricter quality control of niobium }
The niobium sheets for the second series were all eddy-current scanned to eliminate material with tantalum or other foreign inclusions before the deep drawing of half cells. From 715 sheets 637 were found free of defects, 63 showed grinding marks or imprints from rolling and 15 exhibited large signals which in most cases were due to small iron chips. No further Nb sheets with tantalum inclusions were found. Most of the rejected sheets will be recoverable by applying some chemical etching. The iron inclusions were caused by mechanical wear of the rolls used for sheet rolling. In the meantime new rolls have been installed. 
The eddy-current check has turned out to be an important quality control not only for the cavity manufacturer but also for the supplier of the niobium sheets. 

\subsubsection*{(2) Weld preparation}
Stringent requirements were imposed on the electron-beam welding procedure to prevent the degraded performance at the equator welds encountered in the first series. After mechanical trimming the weld regions were requested to be cleaned by a slight chemical etching, ultrapure water rinsing and clean-room drying not more than 8 hours in advance of the EB welding. The success of these two additional quality control measures has been convincing: no foreign material inclusions nor weld contaminations were found in the cavities tested so far.

\subsubsection*{(3) Replacement of Nb flanges by NbTi flanges} 
In the first cavity series the flanges at the beam pipes and the coupler ports were made by rolling over the 2 mm thick niobium pipes.  The sealing against the stainless steel counter flanges was provided by Helicoflex gaskets. This simple design appeared satisfactory in a number of prototype cavities but proved quite unreliable in the series production, mainly due to a softening of the niobium during the 1400$^\circ$C heat treatment. Most of the nine-cell cavities had to be flanged more than once to become leak tight in superfluid helium. This caused not only time delays but also severe problems with contamination and field emission. Therefore an alternative flange design was needed \cite{Zapfe98}. The material was selected to be EB-weldable to niobium and to possess a surface hardness equivalent to that of standard UHV flange material (stainless steel 316 LN/ DIN 4429). Niobium-titanium conforms to these requirements at a reasonable cost. Contrary to pure niobium the alloy NbTi (ratio 45/55 in wt \%) shows no softening after the 1400$^\circ$C heat treatment and only a moderate crystal growth. O-ring type aluminum gaskets provide reliable seals in superfluid helium and are easier to clean than Helicoflex gaskets. During cavity etching the sealing surface must be protected from the acid. 

\subsubsection{Test results in vertical cryostat} 
\begin{figure}
\begin{center}
\epsfig{file=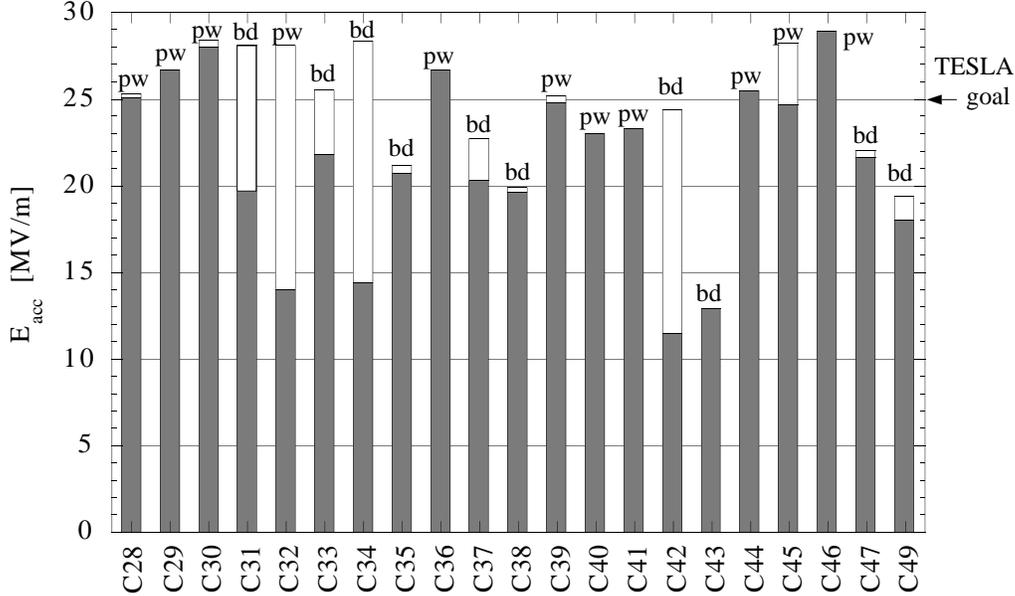,scale=0.8,angle=90}
\caption{\label{F4.12} Test results of the second cavity series (cross-hatched bars); plotted is the highest gradient achieved in the first rf test of each cavity at $Q_0 \ge 5 \cdot 10^9$. Cavities with poor initial performance were subjected to an additional BCP and high pressure water rinsing and tested again (open bars). Field limitation by amplifier power (pw) or thermal breakdown (bd) is indicated for the best gradient.}
\end{center}
\end{figure}

All new cavities were subjected to the standard treatment described in Sect. 4.2, including the post-purification with titanium getter at 1400$^\circ$C. Twenty resonators have been tested up to date. Only one rf test was performed for each resonator in the first round. If some limitation was found the cavity was put aside for further treatment. The results of the first test sequence are summarized in Fig.~\ref{F4.12}. It is seen that 8 cavities reach or exceed the TESLA specification of $E_{acc}\ge \,$ 25 MV/m with a quality factor above $5 \cdot 10^{9}$. Eight cavities are in the range of 18 to 23 MV/m while four cavities show a much lower performance. In cavity C43 a hole was burnt during equator welding which was repaired by welding in a niobium plug; the cavity quenched at 13 MV/m at exactly this position. It is rather unlikely that C43 can be recovered by repeating the repair. Therefore, in future cavity production repaired holes in EB welds will no longer be acceptable.
The cavities C32, C34 and C42 showed very strong field emission in the first test. They have been improved in the meantime by additional BCP and high pressure water rinsing, see Fig.~\ref{F4.12}. Excluding the defective cavity C43, the average gradient is $25.0\pm3.2$ MV/m at $Q_0 = 5 \cdot 10^9$.

\subsubsection{Tests with main power coupler in horizontal cryostat}

\begin{figure}
\begin{center}
\epsfig{figure=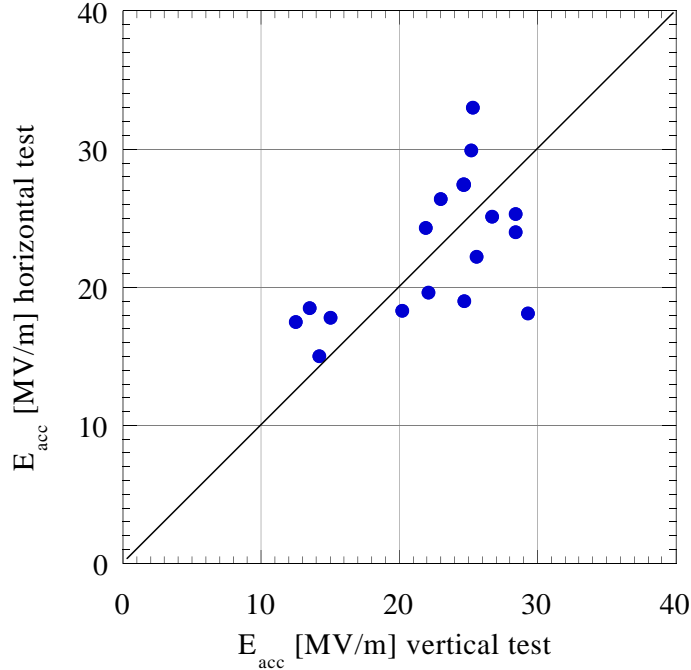,scale=0.7,angle=90}
\caption{\label{F4.13} Comparison of vertical and horizontal test results. The average accelerating field achieved in the vertical test with cw excitation is 22.3 MV/m, in the horizontal test with pulsed excitation 22.5 MV/m. Most of these cavities are from the second production.}
\end{center}
\end{figure}

After the successful test in the vertical bath cryostat the cavities are welded into their liquid helium container and equipped with the main power coupler. The external $Q$ is typically $2\cdot 10^6$, while in the vertical test an input antenna with an external $Q$ of more than $10^{11}$ is used. Four cavities of the first production series and thirteen of the second series have been tested together with their main power coupler in a horizontal cryostat. The accelerating fields achieved in the vertical and the horizontal test are quite similar as shown in Fig. ~\ref{F4.13}. In a few cases reduced performance was seen due to field emission while several cavities improved their field capability due to the fact that with the main power coupler pulsed operation is possible instead of the cw operation in the vertical cryostat. These results indicate that the good performance of the cavities can indeed be preserved after assembly of the liquid helium container and the power coupler provided extreme care is taken to avoid foreign particles from entering the cavity during these assembly steps. 

\subsubsection{Cavity improvement by heat treatment}
\begin{figure}
\begin{center}
\epsfig{figure=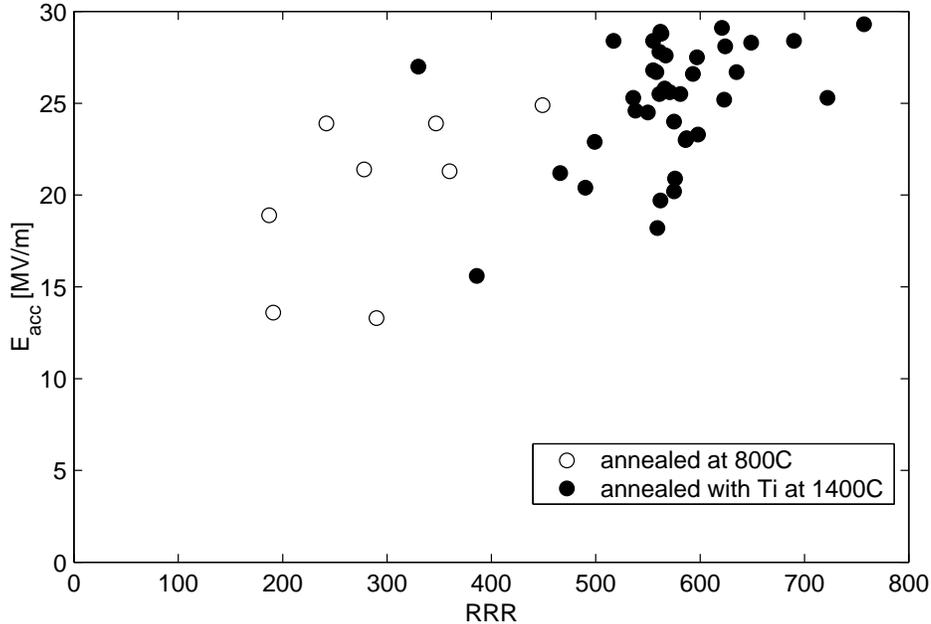,width=12.4cm}
\caption{\label{emax_rrr} Maximum gradient as a function of $RRR$.}
\end{center}
\end{figure}

The beneficial effect of the 800$^\circ$C and 1400$^\circ$C heat treatments has been shown in Fig. 13c. Ten of the 9-cell cavities have been tested after the intermediate 800 $^\circ$C step yielding an average gradient of 20.7 MV/m. The 1400 $^\circ$C treatment with titanium getter raised the average gradient to 24.4 MV/m. It should be noted that part of the 3.7 MV/m improvement may be due to the additional etching of about 80 $\mu$m. An interesting correlation is obtained by plotting the maximum gradient as a function of the measured $RRR$ of the cavity, see Fig.~\ref{emax_rrr}. This figure clearly indicates that a higher heat conductivity leads to higher accelerating fields, at least if the standard BCP treatment is applied to prepare the cavity surface. 

\section{RF Control System and Performance of the Cavities with Electron Beam \label{rfcontrol}}

\subsection{General demands on the rf control system}

The requirements on the stability of the accelerating field in a superconducting acceleration structure are comparable to those in a normal-conducting cavity. However the nature and magnitude of the perturbations to be controlled are rather different. Superconducting cavities possess a very narrow bandwidth  and are therefore highly susceptible to mechanical perturbations. Significant phase and amplitude errors are induced by the resulting frequency variations. Perturbations can be excited by mechanical vibrations (microphonics), changes in helium pressure and level, or Lorentz forces. Slow changes in frequency, on the time scale of minutes or longer, are corrected by a frequency tuner, while faster  changes are counteracted by an amplitude and phase modulation of the incident rf power. 

The demands on amplitude and phase stability of the TESLA Test Facility cavities are driven by the maximum tolerable energy spread in the TTF linac. The design goal is a relative  spread of $\sigma_E/E$ = 2 $\cdot 10^{-3}$ implying a gradient and phase stability in the order of 1 $\cdot 10^{-3}$ and $1.6^\circ$, respectively. For cost reasons up to 32 cavities will be powered by a single klystron. Hence it is not possible to control individual cavities but only the vector sum of the field vectors in these 32 cavities.

One constraint to be observed is that the rf power needed for control should be minimized. The rf control system must also be robust against variations of system parameters such as beam loading and klystron gain. 

\begin{figure}
\begin{center}
\epsfig{figure=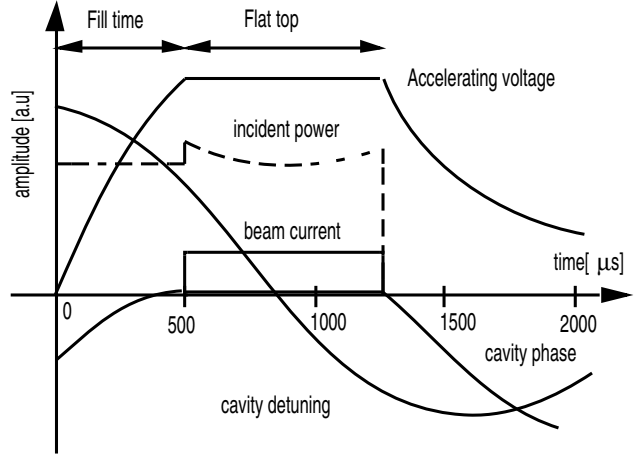,scale=1.15}
\caption{\label{F6.1}
Pulse structure of TTF cavity operation. Shown are: accelerating voltage with 500~$\mu$s filling time and 800~$\mu$s flat top, incident power, beam current, cavity phase and cavity detuning.}
\end{center}
\end{figure}

The pulsed structure of the rf power and the beam at TTF, shown in Fig. \ref{F6.1}, puts demanding requirements on the rf control system. Amplitude and phase control is obviously needed during the flat-top of 800 $\mu$s when the beam is accelerated, but it is equally desirable to control the field during cavity filling to ensure proper beam injection conditions. Field control is aggravated by the transients induced by the subpicosecond electron bunches which have a repetition rate of 1 to 9 MHz.

For a detailed discussion of the basic principles of rf systems used in superconducting electron linacs and their operational performance, refer to \cite{Graef} and \cite{Sim1}.

\subsection{Sources of field perturbations}  

There are two basic mechanisms which influence the magnitude and phase of the accelerating field in a superconducting cavity:
\begin{itemize}
\item
variations in klystron power or beam loading (bunch charge)
\item
modulation of the cavity resonance frequency.
\end{itemize}

Perturbations of the accelerating field through time-varying field excitations are dominated by changes in beam loading. One must distinguish between transients caused by the pulsed structure of the beam current and stochastic fluctuations of the bunch charge. The transients caused by the regular bunch train in the TTF linac (800 picosecond bunches of 8 nC each, spaced by 1 $\mu$s) are in the order of 1\% per 10 $\mu$s; the typical bunch charge fluctuations of 10\% induce field fluctuations of about 1\%. In both cases the effect of the fast source fluctuations on the cavity field is diminuished by the long time constant of the cavity\footnote{The cavity with power coupler is adjusted to an external $Q$ of $3 \cdot 10^6$ at 25 MV/m, corresponding to a time constant of about 700 $\mu$s. The consequence is a low-pass filter characteristic.}. 

\begin{figure}
\begin{center}
\epsfig{figure=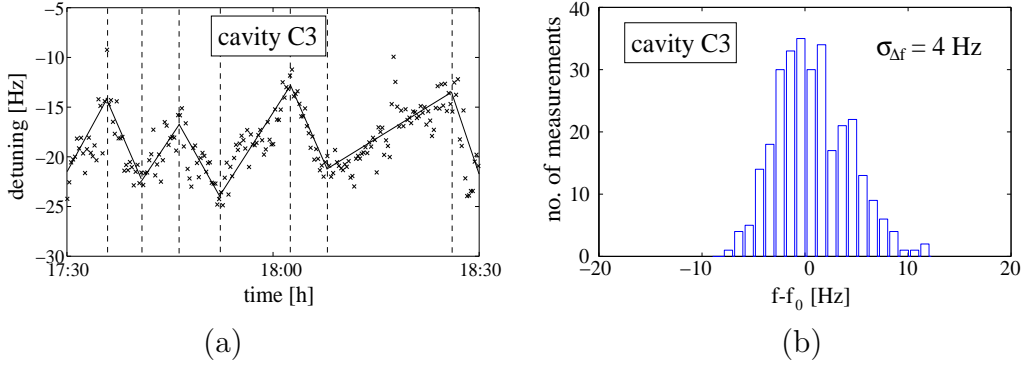,width=13.6cm,height=4.2cm}
$\qquad \qquad \qquad \qquad \qquad \qquad \qquad \qquad \quad$(a) $\qquad \qquad \qquad \qquad \qquad \qquad \qquad \qquad \quad$(b) 
\caption{\label{F6.2}
Fluctuations of  the cavity resonance frequency.
(a) Slow drifts caused by helium pressure variations. The sensitivity is 10 Hz/mbar.
(b) Random variations of resonance frequency after correction for the slow drift.}
\end{center}
\end{figure}

Mechanical changes of the shape and eigenfrequency of the cavities caused by microphonics are a source of amplitude and phase jitter which has bothered superconducting accelerator technology throughout its development. In the TTF cavities the sensitivity of the resonance frequency to a longitudinal deformation is about 300 Hz/$\mu$m. Heavy machinery can transmit vibrations through the ground, the support and the cryostat to the cavity. Vacuum pumps can interact with the cavity through the beam tubes, and the compressors and pumps of the refrigerator may generate mechanical vibrations which travel along the He transfer line into the cryostat. Also helium pressure variations lead to changes in resonance frequency as shown in Fig. \ref{F6.2}a. The rms frequency spread due to microphonics, measured in 16 cavities, is $9.5\pm5.3$~Hz and thus surprisingly small for a superconducting cavity system (see Fig. \ref{F6.2}b). 

\begin{figure}
\begin{center}
\epsfig{figure=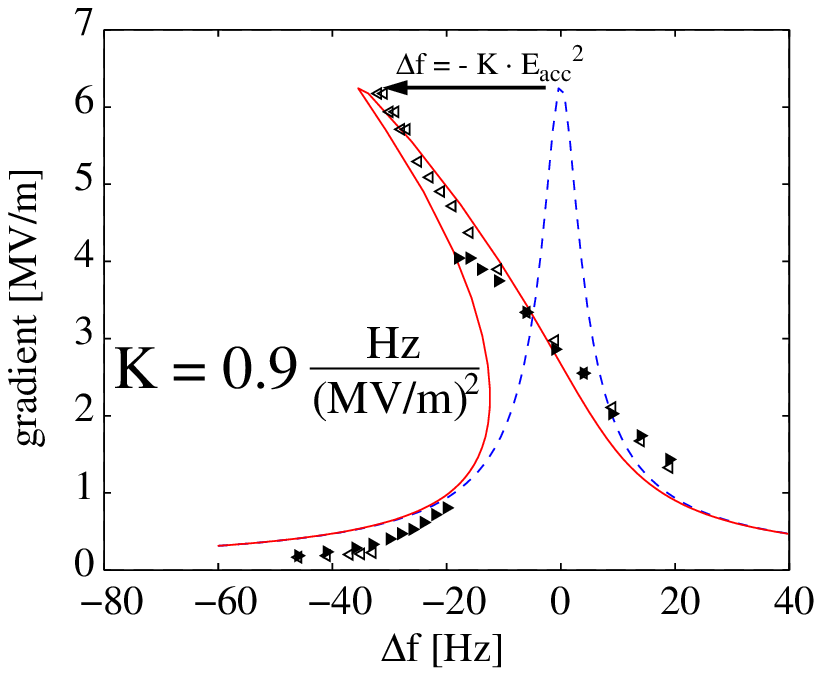,width=6.5cm}
$\quad$ \epsfig{figure=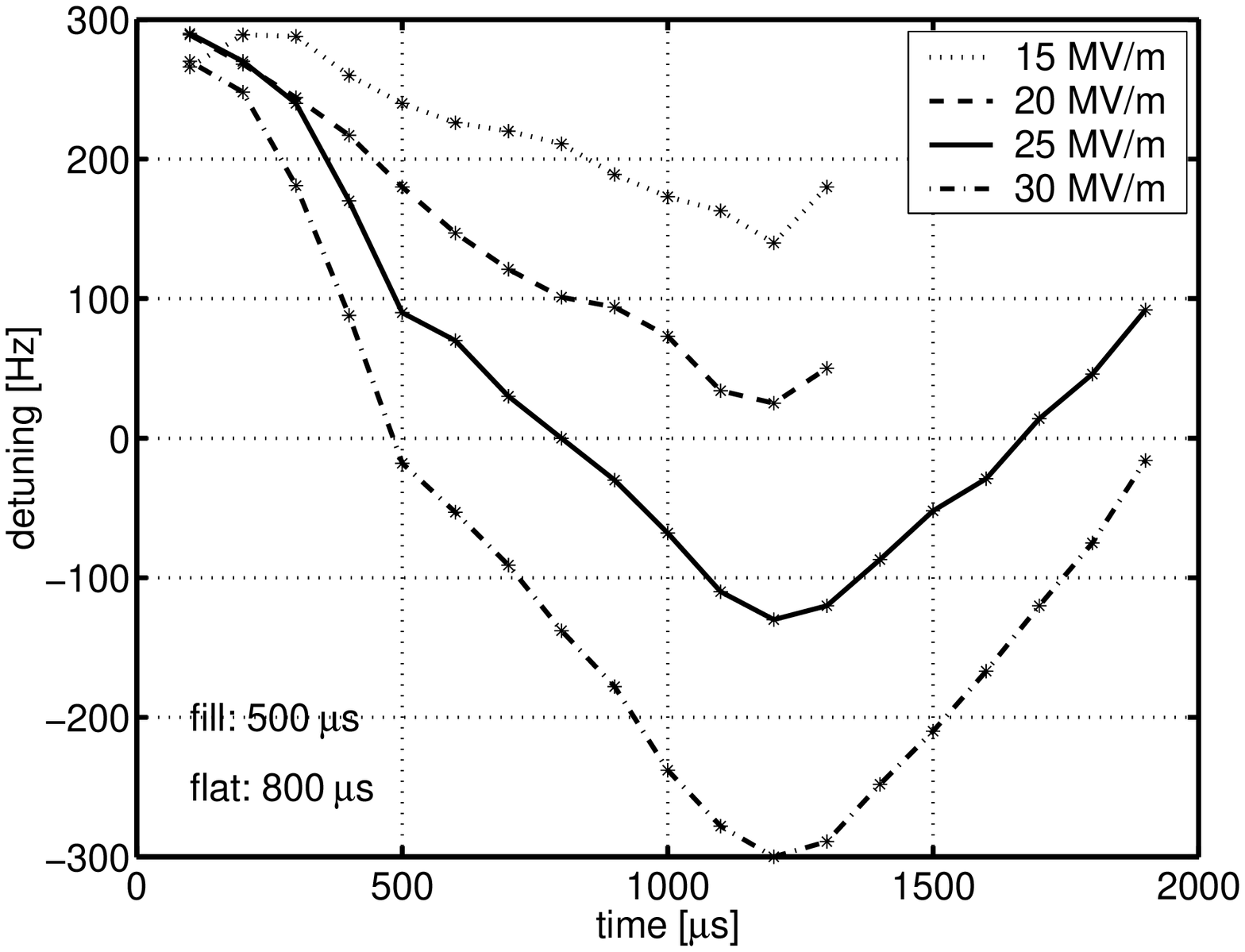,width=6.5cm}
$\qquad \qquad \qquad \qquad \qquad \qquad \qquad \qquad \quad$(a) $\qquad \qquad \qquad \qquad \qquad \qquad \qquad \qquad \quad$(b) 
\caption{\label{F6.3}
(a) Influence of Lorentz forces on the shape of the resonance curve of a sc cavity in cw operation. The left part of the curve was mapped out by approaching the resonance from below, the right part by coming from above. 
 (b) Dynamical detuning of cavity C39 during the TESLA pulse. In pulsed operation the resonance is approached from above.} 
\end{center}
\end{figure}

At high accelerating gradients the Lorentz forces become a severe perturbation. The corresponding frequency shift is proportional to the square of the accelerating field according to $\Delta f = -K \cdot E_{\rm acc}^2$ with 
$K \approx 1$ Hz/(MV/m)$^2$. Figure \ref{F6.3}a shows a cw measurement of the resonance curve with a strong distortion caused by Lorentz forces. In cw operation the frequency shift can be easily corrected for by mechanical tuning. In the pulsed mode employed at the TTF linac this is not possible since the mechanical tuner is far too slow. Hence a time-dependent detuning is unavoidable. In order to keep the deviation from the nominal resonance frequency within acceptable limits the cavities are predetuned before filling. The measured dynamic detuning of cavity C39 during the 1.3 ms long rf pulse is shown in Fig. \ref{F6.3}b for accelerating fields of 15 to 30 MV/m. Choosing a predetuning of $+300$~Hz, the eigenfrequency at 25 MV/m changes dynamically from $+100$~Hz to $-120$~Hz during the 800 $\mu$s duration of the beam pulse.

In steady state (cw) operation at a gradient of 25 MV/m and a beam current of 8 mA a klystron power of 210 kW is required per nine-cell cavity. In pulsed mode $\approx 15\,$ \% additional rf power is needed to maintain a constant accelerating gradient in the presence of cavity detuning. The frequency changes from microphonics and helium pressure fluctuations lead to comparable extra power requirements. The klystron should be operated 10~\% below saturation to guarantee sufficient gain in the feedback loop. 

\subsection{RF control design considerations}

The amplitude and phase errors from Lorentz force detuning, beam transients and microphonics are in the order of 5\% and 20$^\circ$, respectively. These errors must be suppressed by one to two orders of magnitude. Fortunately, the dominant errors are repetitive (Lorentz forces and beam transients) and can be largely eliminated by means of a feedforward compensation. It should be noted, however, that bunch-to-bunch fluctuations of the beam current cannot be suppressed by the rf control system since the gain bandwidth product is limited to about 1 MHz due to the low-pass characteristics of the cavity, the bandwidth limitations of electronics and klystrons, and cable delay. 

\begin{figure}
\begin{center}
\epsfig{figure=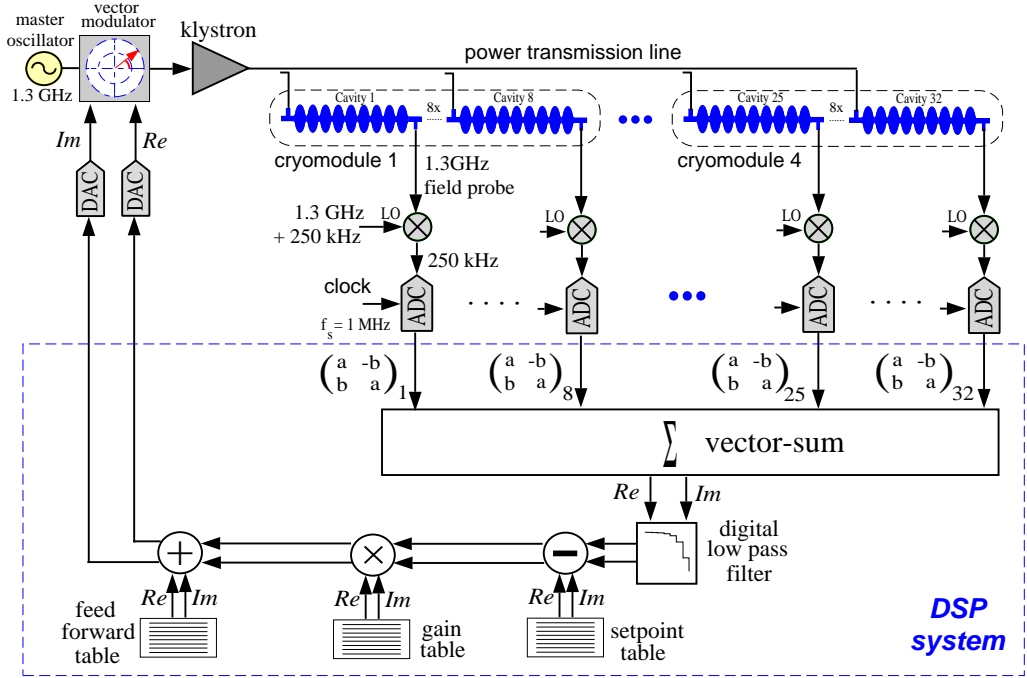,width=13.6cm}
\caption{\label{F6.4}
Schematic of the digital rf control system.} 
\end{center}
\end{figure}

Fast amplitude and phase control can be accomplished by modulation of the incident rf wave which is common to 32 cavities. The control of an individual cavity field is not possible. The layout of the TTF digital rf control system is shown in Fig. \ref{F6.4}. The vector modulator for the incident wave is designed as a so-called ``I/Q modulator'' controlling real and imaginary part of the complex cavity field vector instead of amplitude and phase. This has the advantage that the coupling between the two feedback loops is minimized and the problem of large phase uncertainties at small amplitude is avoided.

The detectors for cavity field, incident and reflected wave are implemented as digital detectors for real and imaginary part. The rf signals are converted to an intermediate frequency of 250 kHz and sampled at a rate of 1 MHz, which means that two subsequent data points yield the real and the imaginary part of the cavity field vectors. These vectors are multiplied with 2$\times$2 rotation matrices to correct for phase offsets and to calibrate the gradients of the individual cavity probe signals. The vector sum is calculated and a Kalman filter is applied which provides an optimal state (cavity field) estimate by correcting for delays in the feedback loop and by taking stochastic sensor and process noise into account. Finally the nominal set point is subtracted and a time-optimal gain matrix is applied to calculate the new actuator setting (the {\it Re} and {\it Im} control inputs to the vector modulator). Adaptive feedforward is realized by means of a table containing the systematic variations, thereby reducing the task of the feedback loop to control the remaining stochastic fluctuations. The feedforward tables are continuously updated to take care of slow changes in parameters such as average detuning angle, microphonic noise level and phase shift in the feedforward path. 

\subsection{Operational experience}

The major purpose of the TESLA Test Facility linac is to demonstrate that all major accelerator subsystems meet the technical and operational requirements of the TESLA collider. Currently the TTF linac is equipped with two cryomodules each containing 8 cavities. The cavities are routinely operated at the design gradient of TTF of 15 MV/m, providing a beam energy of 260 MeV. 
\begin{figure}
\begin{center}
\epsfig{figure=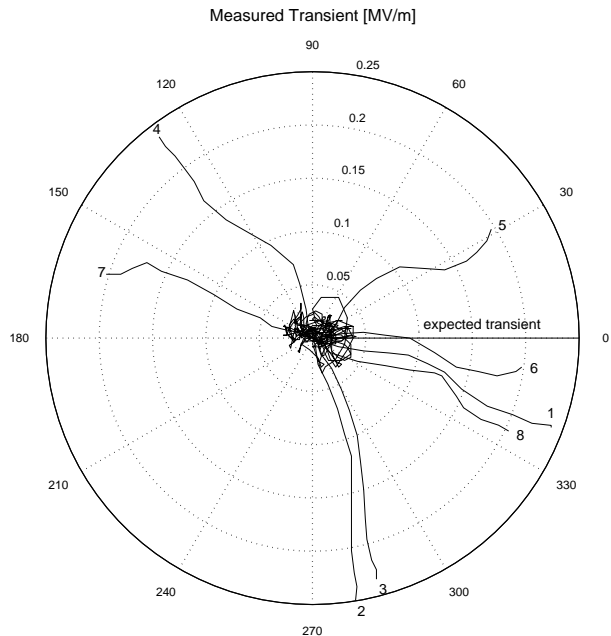,scale=0.45}
\caption{\label{F6.5}
Beam induced transients (cavity field vectors) obtained by measuring the cavity fields with and without beam pulses. The noise in the signals can be estimated from the erratic motion at the center of the plot. This region represents the first  50 $\mu$s of the measurement before the arrival of the beam pulse. }   
\end{center}
\end{figure}

An important prerequisite of the proper functioning of the vector-sum control of 16 to 32 cavities is an equal response of the field pickup probes in the individual cavities. A first step is to adjust the phase of the incident rf wave to the same value in all cavities by means of three-stub tuners in the wave guides. Secondly, the transients induced by the bunched beam are used to obtain a relative calibration of the pickup probes, both in terms of amplitude and phase. Typical data taken at the initial start-up of a linac run are shown in Fig. \ref{F6.5}. Ideally the lengths of the field vectors should all be identical since the signals are induced by the same electron bunch in all cavities. The observed length differences indicate a variation in the coupling of the pickup antenna to the beam-induced cavity field, which has to be corrected. The different phase angles of the field vectors are mainly due to different signal delays. The complex field vectors are rotated by matrix multiplication in digital signal processors to yield all zero phase. Moreover they are normalized to the same amplitude to correct for the different couplings of the pickup antennas to the cavity fields. Once this calibration has been performed the vector sum of the 16 or 32 cavities is a meaningful measure of the total accelerating voltage supplied to the beam. The calibration is verified with a measurement of the beam energy in a magnetic spectrometer. 

\begin{figure}
\begin{center}
\epsfig{figure=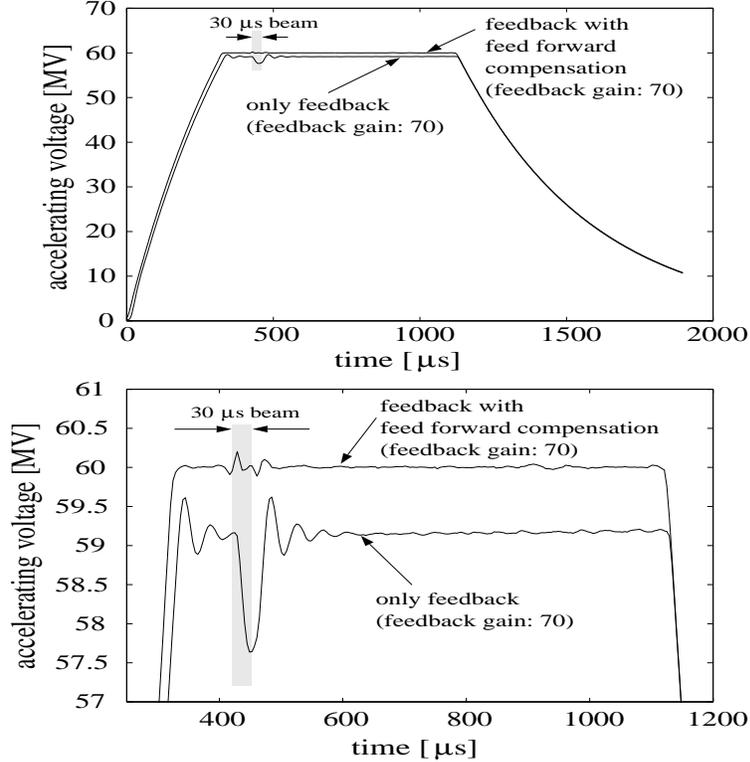,scale=0.65}
\caption{\label{F6.6}
Field regulation of the vector-sum of 8 cavities without and with adaptive feedforward. The lower graph shows an enlarged view of the plateau region.}
\end{center}
\end{figure}

The required amplitude stability of $ 1 \cdot 10^{-3}$ and phase stability of $\sigma_{\phi} \leq 1.6^\circ$ can be achieved during most of the beam pulse duration with the exception of the beam transient induced when turning on the beam. Without feedback the transient of a 30~$\mu$s beam pulse at 8~mA would be of the order of 1~MV/m. This transient can be reduced to about 0.2~MV/m by turning on feedback. The effectiveness of the feedback system is limited by the loop delay of 5~$\mu$s and the unity-gain bandwidth of about 20 kHz. The 0.2~MV/m transient is repetitive with a high degree of reproducibility. Using feedforward it can be further suppressed by more than an order of magnitude, as shown in Fig. \ref{F6.6}. Slow drifts are corrected for by making the feedforward system adaptive \cite{Li_Si}. The feedforward tables are updated on a time scale of minutes. 

\section{Cavities of Higher Gradients \label{highgrad}}

Both the TESLA collider and the X-ray FEL would profit from the development of cavities which can reach higher accelerating fields (i.e. higher particle energies) and higher quality factors (i.e. reduced operating costs of the accelerators).
The TESLA design energy of 250 GeV per beam requires a gradient of 25 MV/m in the present nine-cell cavities. The results shown in Sect. 5.4  demonstrate that TESLA could indeed be realized with a moderate improvement in the present cavity fabrication and preparation methods. However, for particle physics an energy upgrade of the collider would be of highest interest, and hence there is a strong motivation to push the field capability of the cavities closer to the physical limit of about 50 MV/m which is determined by the superheating field of niobium. Three main reasons are known why the theoretical limit has not yet been attained in multicell resonators:
(1) foreign material contamination in the niobium,
(2) insufficient quality and cleanliness of the inner rf surface,
(3) insufficient mechanical stability of the resonators.
An  R\&D program has been initiated aiming at improvements in all three directions. Furthermore, the feasibility of seamless cavities is being investigated. 

\subsection{Quality improvement of niobium}

Niobium for microwave resonators has to be of high purity for several reasons: (a) dissolved gases like hydrogen, oxygen and nitrogen reduce the heat conductivity at liquid helium temperature and degrade the cooling of the rf surface; (b) contamination by foreign metals may lead to magnetic flux pinning and heat dissipation in rf fields; (c) normal-conducting or weakly superconducting clusters close to the rf surface are particularly dangerous . The Nb ingots contain about 500 ppm of finely dispersed tantalum. It appears unlikely that the Ta clusters found in some early TTF cavities might have been caused by this ``natural'' Ta content. Rather there is some suspicion that Ta grains might have dropped into the Nb melt during the various remeltings of the Nb ingot in an electron-beam melting furnace because such furnaces are often used for Ta production as well. 
To avoid contamination by foreign metals a dedicated electron-beam melting furnace would appear highly desirable but seems to be too cost-intensive in the present R\&D phase of TESLA. Also more stringent requirements on the quality of the furnace vacuum (lower pressure, absence of hydrocarbons) would improve the Nb purity. The production steps following the EB melting (machining, forging and sheet rolling of the ingot) may also introduce dirt. The corresponding facilities need careful inspection and probably some upgrading.
The present TTF cavities have been made from niobium with gas contents in the few ppm range and an $RRR$ of 300. Ten 9-cell cavities have been measured both after 800$^\circ$C and 1400$^\circ$C firing. The average gain in gradient was about 4 MV/m. It would be highly desirable to eliminate the tedious and costly 1400$^\circ$C heat treatment of complete cavities. One possibility might be to produce a niobium ingot with an $RRR$ of more than 500. This is presently not our favored approach, mainly for cost reasons. 

For the present R\&D program, the main emphasis is on the production of ingots with $RRR\ge 300$, but with improved quality by starting from niobium raw material with reduced foreign material content, especially tantalum well below 500 ppm. Stricter quality assurance during machining, forging and sheet rolling should prevent metal flakes or other foreign material from being pressed into the niobium surface deeper than a few $\mu$m.
To increase the $RRR$ from 300 to about 600, it is planned to study the technical feasibility\footnote{At Cornell University cavities have been successfully fabricated from $RRR=1000$ material. Likewise, the TTF cavity C19 was made from post-purified half cells and showed good performance.} of a 1400$^\circ$C heat treatment at the dumb-bell stage (2 half cells joined by a weld at the iris). This procedure would be preferable compared to the heat treatment of whole cavities which must be carefully supported in a Nb frame to prevent plastic deformation, while such a precaution is not needed for dumb-bells. However, there is a strong incentive to find cavity treatment methods which would permit elimination of  the 1400$^\circ$C heat treatment altogether. According to the results obtained at KEK \cite{saito} electropolishing seems to offer this chance (see below).

\subsection{Improvement in cavity fabrication and preparation}

Once half cells or dumb-bells of high $RRR$ have been produced it is then mandatory to perform the electron-beam welding of the cavities in a vacuum of a few times $10^{-6}$ mbar in order to avoid degradation of the $RRR$ in the welds. The EB welding machines available at industrial companies achieve vacua of only $5 \cdot 10^{-5}$ mbar and are hence inadequate for this purpose. An EB welding machine at CERN is equipped with a much better vacuum system. This EB apparatus is being used for a single-cell test program. For the future cavity improvement program a new electron-beam welding apparatus will be installed at DESY with a state-of-the-art electron gun, allowing computer-controlled beam manipulations, and with an oilfree vacuum chamber fulfilling UHV standards. 

The industrially produced cavities undergo an elaborate treatment at TTF before they can be installed in the accelerator. A 150--200 $\mu$m thick damage layer is removed from the rf surface because otherwise gradients of 25 MV/m appear inaccessible. As explained in Sect.~4 the present method is Buffered Chemical Polishing (BCP) which leads to a rather rough surface with strong etching in the grain boundaries. An alternative method is ``electropolishing'' (EP) in which the material is removed in an acid mixture under current flow. Sharp edges and burrs are smoothed out and a very glossy surface can be obtained. For a number of years remarkable results have been obtained at KEK with electropolishing of 1-cell niobium cavities. Recently, a collaboration between KEK and Saclay has convincingly demonstrated that EP raises the accelerating field by more than 7 MV/m with respect to BCP. Several 1-cell cavities from Saclay, which showed already good performance after the standard BCP, exhibited a clear gain after the application of EP \cite{kako}. Conversely, an electropolished cavity which had reached 37 MV/m suffered a degradation after subsequent BCP. These results are a strong indication that electropolishing is the superior treatment method. 

CERN, DESY, KEK and Saclay started a joint R\&D program with electropolishing of half cells and 1-cell cavities in August 1998. Recent test results yield gradients around 40 MV/m \cite{Lilje99} and hence the same good performance as was achieved at KEK. The transfer of the EP technology to 9-cell resonators requires considerable effort. It is planned to do this in collaboration with industry.

\begin{figure}
\begin{center}
\epsfig{figure=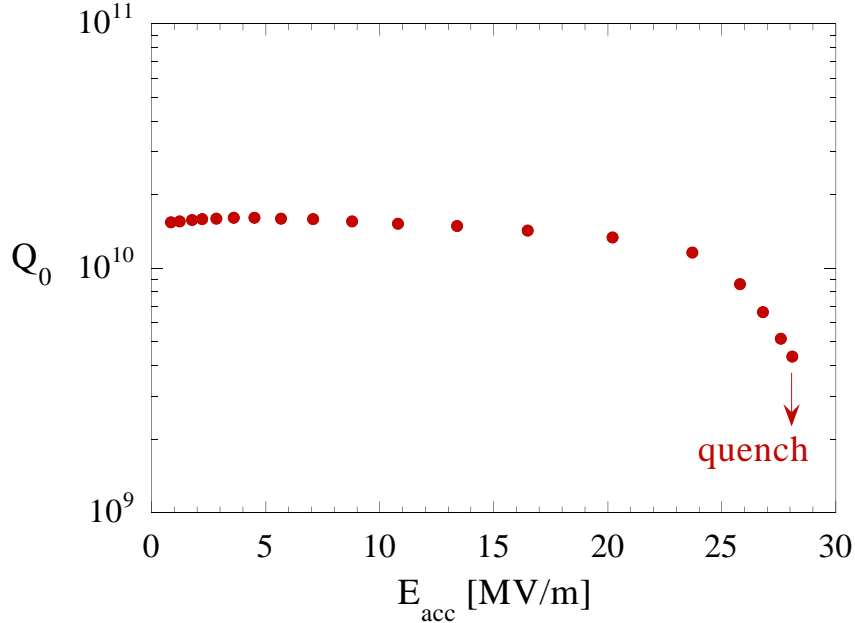,scale=0.60,angle=90}
\caption{\label{S28X} Excitation curve of a TESLA 9-cell cavity showing a drop of quality factor without any field emission.}
\end{center}
\end{figure}

Recently it has been found \cite{pkneis} that an essential prerequisite for achieving gradients in the 40 MV/m regime is a baking at 100 to 150$^\circ$C for up to 48 hours while the cavity is evacuated, after the final high-pressure water rinsing. In electropolished cavities this procedure removes the drop of quality factor towards high gradients which is often observed without any indication of field emission. Such a drop is usually also found in chemically etched cavities; see for example Fig. \ref{S28X}. Experiments at Saclay \cite{saclay} have shown that a baking may improve the $Q(E)$ curve; however, part of the $Q$ reduction at high field may be due to local magnetic field enhancements at the sharp grain boundaries of BCP treated cavities \cite{knob}. 

\subsection{Mechanical stability of the cavities}

The stiffening rings joining neighboring cells in the TESLA resonator are adequate to limit Lorentz-force detuning up to accelerating fields of 25 MV/m. Beyond 25 MV/m the cavity reinforcement provided by these rings is insufficient. Hence an alternative stiffening scheme must be developed for cavities in the 35--40 MV/m regime. 
A promising approach has been taken at Orsay and Saclay. The basic idea is to reinforce a thin-walled niobium cavity by a 2 mm thick copper layer which is plasma-sprayed onto the outer wall. Several successful tests have been made \cite{orsay}. 
The copper plating has a potential danger since Nb and Cu have rather different thermal contraction. The deformation of a cavity upon cooldown and the resulting frequency shift need investigation. Another phenomenon has been observed in cavities made from explosion-bonded niobium-copper sheets: when these cavities were quenched, a reduction in quality factor $Q_0$ was observed \cite{saito2}. An explanation maybe trapped magnetic flux from thermo-electric currents at the copper-niobium interface. It is unknown whether this undesirable effect happens also in copper-sprayed cavities. 
An alternative to copper spraying might be the reinforcement of a niobium cavity by depositing some sort of metallic ``foam'', using the plasma or high velocity spraying technique. If the layer is porous the superfluid helium penetrating the voids should provide ample cooling. 

The cavity reinforcement by plasma or high-velocity spraying appears to be a promising approach but considerable R\&D work needs to be done to decide whether this is a viable technique for the TESLA cavities. 

\subsection{Seamless cavities}

The EB welds in the present resonator design are a potential risk. Great care has to be taken to avoid holes, craters or contamination in the welds which usually have a detrimental effect on the high-field capability. A cavity without weld connections in the regions of high electric or magnetic rf field would certainly be less vulnerable to small mistakes during fabrication. For this reason the TESLA collaboration decided several years ago to investigate the feasibility of producing seamless cavities. Two routes have been followed: spinning and hydroforming.

At the Legnaro National Laboratory of INFN in Italy the spinning technique \cite{palmieri} has been successfully applied to form cavities out of niobium sheets. The next step will be to produce a larger quantity of 1-cell, 3-cell and finally 9-cell cavities from seamless Nb tubes with an $RRR$ of 300. In the cavities spun from flat sheets a very high degree of material deformation was needed, leading to a rough inner cavity surface. Gradients between 25 and 32 MV/m were obtained after grinding and heavy etching (removal of more than 500 $\mu$m) \cite{knei_palm}. Starting from a tube the amount of deformation will be much less and a smoother inner surface can be expected. This R\&D effort is well underway and the first resonators can be expected in early 2000.

The hydroforming of cavities from seamless niobium tubes is being pursued at DESY \cite{gonin}. Despite initial hydroforming difficulties, related to inhomogeneous mechanical properties of the niobium tubes, four single cell cavities have been successfully built so far. Three of these were tested and reached accelerating fields of 23 to 27 MV/m. In a very recent test\footnote{Preparation and test of the hydroformed cavities were carried out by P. Kneisel at Jefferson Laboratory, Newport News, USA.} 32.5 MV/m was achieved at $Q_0=2 \cdot 10^{10}$. Most remarkable is the fact that the cavity was produced from low $RRR$ niobium ($RRR=100$). It received a 1400$^{\circ}$C heat treatment raising the $RRR$ to 300--400. The surface was prepared by grinding and 250 $\mu$m BCP. 

\subsection{Niobium sputtered cavities}
Recent investigations at CERN \cite{benvenuti2} and Saclay \cite{bosland} show that single-cell copper cavities with a niobium sputter layer of about 1~$\mu$m thickness are able to reach accelerating fields beyond 20 MV/m. These results appear so promising that CERN and DESY have agreed to initiate an R\&D program on 1.3 GHz single cell sputtered cavities aiming at gradients in the 30 MV/m regime and quality factors above $5\cdot 10^9$. High-performance sputtered cavities would certainly be of utmost interest for the TESLA project for cost reasons. Another advantage would be the suppression of Lorentz force detuning by choosing a sufficiently thick copper wall.

\subsection{The superstructure concept}

The present TTF cavities are equipped with one main power coupler and two higher order mode couplers per 9-cell resonator. The length of the interconnection between two cavities has been set to $3\lambda/2$ ($\lambda=0.23$ m is the rf wavelength) in order to suppress cavity-to-cavity coupling of the accelerating mode. A shortening of the interconnection is made possible by the ``superstructure'' concept, devised by J.~Sekutowicz \cite{Sek}. Four 7-cell cavities of TESLA geometry are joined by beam pipes of length ($\lambda/2$). The pipe diameter is increased to permit an energy flow from one cavity to the next, hence one main power coupler is sufficient to feed the entire superstructure. One HOM coupler per short beam pipe section provides sufficient damping of dangerous higher modes in both neighboring cavities.
Each 7-cell cavity will be equipped with its own LHe vessel and frequency tuner. Therefore, in the superstructure the field homogeneity tuning (equal field amplitude in all cells) and the HOM damping can be handled at the sub-unit level. The main advantages of the superstructure are an increase in the active acceleration length in TESLA - the design energy of 250 GeV per beam can be reached with a gradient of 22 MV/m - and a savings in rf components, especially power couplers.

A copper model of the superstructure is presently used to verify the theoretically predicted performance. This model allows individual cell tuning, field profile adjustment, investigation of transients in selected cells, test of the HOM damping scheme and measurement of the cavity couplings to the fundamental mode coupler. Also the influence of mechanical tolerances is studied. First results are promising \cite{Csfe}. 
A niobium superstructure prototype is under construction and will be tested with beam in the TTF linac beginning of 2001.

\section{Acknowledgements}

We are deeply indebted to the late distinguished particle and accelerator physicist Bj{\o}rn H. Wiik who has been the driving force behind the TESLA project and whose determination and enthusiasm was essential for much of the progress that has been achieved. We want to thank the many physicists, engineers and technicians in the laboratories of the TESLA collaboration and in the involved industrial companies for their excellent work and support of the superconducting cavity program. We are grateful to P. Kneisel for numerous discussions.

\end{document}